\documentclass[11pt]{article}
\usepackage{times}
\usepackage{a4wide}
\usepackage{latexsym}
\usepackage{amsmath}
\usepackage{amssymb}
\usepackage{ifthen}
\usepackage{mathrsfs}
\usepackage{graphics}
\usepackage{color}
\usepackage{tikz}
\usepackage{stmaryrd}
\usepackage{amsthm}

\newcommand{\nc}{\newcommand}
\nc{\rnc}{\renewcommand} \nc{\nev}{\newenvironment}
\rnc{\subsection}{\secdef\ssa\ssb}
\nc{\ssa}[2][default]{\par\vspace{1ex}\refstepcounter{subsection}\noindent\textbf{\thesubsection.
#1. }} \nc{\ssb}[1]{\par\vspace{2ex}\noindent\textbf{#1. }}

\rnc{\subsubsection}{\secdef\sssa\sssb}
\nc{\sssa}[2][default]{\par\vspace{1ex}\refstepcounter{subsubsection}\noindent\textit{\thesubsubsection.
#1. }} \nc{\sssb}[1]{\par\vspace{1ex}\noindent\textit{#1. }}

\makeatletter
\rnc{\@seccntformat}[1]{{\normalfont\bfseries{\csname
the#1\endcsname}\hspace{1pt}.\hspace{0.4em}}}
\rnc{\section}{\@startsection
        {section}%
        {1}%
        {0mm}%
        {-\baselineskip}%
        {0.5\baselineskip}%
        {\normalfont\normalsize\bfseries\centering}%
}
\renewcommand{\@makecaption}[2]{\begin{center}#1. #2\end{center}}
\makeatother

\newtheorem{theo}{Theorem}[section]
\newtheorem{lem}[theo]{Lemma}
\newtheorem{cor}[theo]{Corollary}
\newtheorem{prop}[theo]{Proposition}

\theoremstyle{definition}
\newtheorem{defn}[theo]{Definition}
\newtheorem{rem}[theo]{Remark}
\newtheorem{exa}[theo]{Example}

\rnc{\proof}[1][{}]{\smallskip\noindent\textit{Proof #1: }}
\nc{\proofend}{\hfill$\Box$\vspace{\topsep}\par}

\rnc{\labelenumi}{(\arabic{enumi})} \rnc{\labelitemi}{\text{--}}
\rnc{\phi}{\varphi} \rnc{\epsilon}{\varepsilon}
\nc{\bigmid}{\;\big|\;} \nc{\Bigmid}{\;\Big|\;}
\rnc{\max}{\textup{max}} \rnc{\min}{\textup{min}}
\rnc{\log}{\textup{log}\;}

\newlength{\probwidth}
\nc{\prob}[3][9]{
\begin{center}
  \normalfont\fbox{
   \begin{tabular}[t]{
     rp{#1cm}}\textit{Instance:}&#2. \\
     \textit{Problem:}&#3
   \end{tabular}}
\end{center}}

\nc{\pprob}[4][9]{
\begin{center}
   \normalfont\fbox{
    \begin{tabular}[t]{
     rp{#1cm}}\textit{Instance:}&#2. \\
     \textit{Parameter:}&#3. \\
     \textit{Problem:}&#4
   \end{tabular}}
\end{center}}

\nc{\nprob}[4][9]{
\begin{center}
  \normalfont\fbox{

\addtolength{\probwidth}{#1cm}\parbox{\probwidth}{\textsc{#2}\\\hspace*{1.5em}
     \begin{tabular}[t]{
      rp{#1cm}}\textit{Instance:}&#3. \\
      \textit{Problem:}&#4
     \end{tabular}}}
\end{center}}

\nc{\npprob}[5][9]{
\begin{center}
  \normalfont\fbox{

\addtolength{\probwidth}{#1cm}\parbox{\probwidth}{\textsc{#2}\\\hspace*{1.5em}
    \begin{tabular}[t]{
     rp{#1cm}}\textit{Instance:}&#3. \\
     \textit{Parameter:}&#4. \\
     \textit{Question:}&#5
    \end{tabular}}}
\end{center}}

\nc{\nppxrob}[5][9]{ \normalfont\fbox{

\addtolength{\probwidth}{#1cm}\parbox{\probwidth}{\textsc{#2}\\\hspace*{1.5em}
   \begin{tabular}[t]{
    rp{#1cm}}\textit{Instance:}&#3. \\
    \textit{Parameter:}&#4. \\
    \textit{Problem:}&#5
   \end{tabular}}}}

\nc{\nppprob}[5][4]{
\begin{center}
  \normalfont\fbox{

\addtolength{\probwidth}{#1cm}\parbox{\probwidth}{\textsc{#2}\\\hspace*{1.5em}
    \begin{tabular}[t]{
     rp{#1cm}}\textit{Instance:}&#3. \\
     \textit{Parameter:}&#4. \\
     \textit{Problem:}&#5
    \end{tabular}}}
\end{center}}

\nc{\noptprob}[6][9]{
\begin{center}
  \normalfont\fbox{

\addtolength{\probwidth}{#1cm}\parbox{\probwidth}{\textsc{#2}\\\hspace*{1.5em}
    \begin{tabular}[t]{
     rp{#1cm}}\textit{Instance:}&#3. \\
     \textit{Solution:}&#4. \\
     \textit{Cost:}&#5. \\
     \textit{Goal:}&#6.
    \end{tabular}}}
\end{center}}

\nc{\FOR}{\textbf{for}}
\nc{\FORALL}{\textbf{for all}}
\nc{\TO}{\textbf{to}}
\nc{\DO}{\textbf{do}}
\nc{\OD}{\textbf{od}}
\nc{\IF}{\textbf{if}}
\nc{\FI}{\textbf{fi}}
\nc{\THEN}{\textbf{then}}
\nc{\ELSE}{\textbf{else}}
\nc{\WHILE}{\textbf{while}}
\nc{\REPEAT}{\textbf{repeat}}
\nc{\UNTIL}{\textbf{until}}
\nc{\OR}{\textbf{or}}
\nc{\AND}{\textbf{and}}
\nc{\PRINT}{\textbf{print}}

\nc{\im}[1]{\item\hspace{#1cm}}
\nev{algorithm}{\begin{enumerate}\rnc{\labelenumi}{\textit{\small \arabic{enumi}.}}\rnc{\itemsep}{0ex}}{\end{enumerate}}

\nc{\fpcl}[1]{\left[#1\right]_{\text{\upshape fp}}}
\nc{\pr}{\le^{\text{\normalfont fp}}_m} \nc{\FPT}{\textup{FPT}}
\nc{\EPT}{\textup{EPT}} \nc{\SUBEPT}{\textup{SUBEPT}}

\nc{\fpt}{\textup{fpt}} \nc{\fptT}{\textup{fpt-T}}
\nc{\W}[1]{\text{$\textup{W}[#1]$}}
\nc{\M}[1]{\text{$\textup{M}[#1]$}}
\nc{\MS}[2]{\text{$\textup{M}^{#1}[#2]$}}
\nc{\MINI}[1]{\mbox{\small \rm MINI[$#1$]}}
\nc{\WP}{\textup{W[P]}} \nc{\AWP}{\textup{AW[P]}}

\rnc{\S}[1]{\text{$\textup{S}[#1]$}} \nc{\SP}{\textup{S[P]}}
\nc{\MP}{\textup{M[P]}}

\nc{\PTIME}{\textup{PTIME}} \nc{\APTIME}{\textup{APTIME}}
\nc{\PSPACE}{\textup{PSPACE}} \nc{\NP}{\textup{NP}}

\nc{\DTIME}{\textup{DTIME}}

\nc{\se}{\subseteq} \nc{\re}{\rightarrow}

\nc{\LOEFF}[1]{{o}^{\rm eff}(#1)}

\nc{\PNPTC}{\mbox{$\textup{P}[{\textsc{tc}}]\ne
\textup{NP}[{\textsc{tc}}]$}} \nc{\str}[1]{\ensuremath{\mathcal #1}}
\nc{\cls}[1]{\ensuremath{\mathbf #1}}

\nc{\algo}[1]{{\mathbb #1}}

\nc{\NAT}{{\mathbb N}}

\nc{\VC}{\textsc{Vertex-Cover}} \nc{\IS}{\textsc{Independent-Set}}
\nc{\Cli}{\textsc{Clique}} \nc{\DS}{\textsc{Dominating-Set}}
\nc{\TSAT}{\textsc{3-Sat}}

\nc{\CNF}{\textup{CNF}} \nc{\WSAT}{\textsc{WSat}}
\nc{\AWSAT}{\textsc{AWSat}} \nc{\SAT}{\textsc{Sat}}
\nc{\ASAT}{\textsc{ASat}} \nc{\CIRC}{\textsc{Circ}}
\nc{\PROP}{\textsc{Prop}}

\nc{\nva}{\textup{nv}} \nc{\ncl}{\textup{nc}}

\nc{\ETH}{\textup{ETH}}
\nc{\Wone}{\textup{W[1]}}

\nc{\serf}{\textup{serf}} \nc{\serfT}{\textup{serf-T}}
\nc{\var}{\textup{var}}

\nc{\NUXP}{\textup{XP}_{\rm nu}} \nc{\XP}{\textup{XP}}
\nc{\SNP}{\textup{SNP}}
\nc{\X}{\textup{X}}

\nc{\ept}{{\rm ept}}

\nc{\E}{\textup{E}}

\nc{\HALT}{\textsc{Halt}}

\nc{\TM}{\textsc{TM}} \nc{\TMBA}{\textsc{TMBA}}

\nc{\ceil}[1]{\left\lceil#1\right\rceil}
\nc{\floor}[1]{\left\lfloor#1\right\rfloor}

\nc{\bende}{\eqno$\Box$} \nc{\benda}{\tag*{$\Box$}}

\nc{\pa}{\kappa}

\nc{\co}{\textup{co-}}

\rnc{\L}{\textup{LOGSPACE}}

\nc{\NL}{\textup{NLOGSPACE}}

\rnc{\P}{\textup{P}}

\rnc{\angle}[1]{\langle #1\rangle}

\nc{\rand}[1]{\marginpar{\raggedright\footnotesize #1}}
\nc{\yrand}[1]{\rand{\textbf{Y: }#1}}
\nc{\jrand}[1]{\rand{\textbf{J: }#1}}
\nc{\rjrand}[1]{\rand{\textcolor{red}{\textbf{J: }#1}}}

\nc{\Ppoly}{\textup{P}/\text{\small poly}}

\nc{\f}{\mathbf f}

\nc{\s}{\mathbf s}

\nc{\tim}{\textup{time}}

\nc{\sat}{\textup{sat}}
\nc{\rank}{\textup{rank}}
\nc{\PC}{\mathbf{PC}}
\nc{\bin}{{\rm in}}
\nc{\bout}{{\rm out}}
\nc{\Mix}{\textsc{Mix}}
\nc{\MIX}{\mathbf{MIX}}

\nc{\tdeg}{\textup{tdeg}}
\nc{\lspan}{\textup{span}}

\nc{\ma}[1]{\mathbb #1}
\nc{\bet}[1]{\| #1\|}
\nc{\EFM}[2]{(#1,#2)}
\nc{\EF}{\text{Ehren\-feucht\--Fra\"i\-ss\'e}}
\nc{\AF}{\textup{AF}}
\nc{\supp}{\textup{supp}}
\nc{\ar}{\textup{ar}}
\nc{\pol}{\textup{pol-}}
\nc{\equivm}{\equiv_{\LFP_m}}
\nc{\equivfo}{\equiv_{\FO_m}}
\nc{\ERRE}{Erd\H{o}s-R\'enyi}

\nc{\PCC}{\textup{PCC}}

\nc{\TCOL}{\textsc{3-Col}}

\nc{\GI}{\textup{GI}}

\nc{\EX}{\textup{E}}

\nc{\Var}{\textup{Var}}

\newcommand{\AC}{\textup{AC}}
\newcommand{\ER}{\textup{ER}}

\nc{\size}{\textup{size}}

\nc{\DNF}{\textup{DNF}}

\nc{\n}{\tilde n}

\nc{\dotcup}{\;\dot\cup\;}

\nc{\ds}{\gamma}

\nc{\mds}{\ensuremath{\textsc{Min-Dominating-Set}}}
\nc{\pds}{\ensuremath{p\textsc{-Dominating-Set}}}
\nc{\pclique}{\ensuremath{p\textsc{-Clique}}}
\nc{\pvc}{\ensuremath{p\textsc{-Vertex-Cover}}}

\nc{\pwsat}{\ensuremath{p\textsc{-WSat}}}

\nc{\mcs}{\ensuremath{\textsc{Monotone-Circuit-Satisfiability}}}

\nc{\vc}{\textsc{Vertex-Cover}}
\nc{\clique}{\textsc{Clique}}

\nc{\sol}{\textup{sol}}
\nc{\cost}{\textup{cost}}
\nc{\goal}{\textup{goal}}

\nc{\pow}{\text{Pow}}

\nc{\mwds}{\textsc{Min-Weighted-Dominating-Set}}

\nc{\paraAC}{\textup{para-$\AC^0$}}

\nc{\para}{\textup{para-}}

\nc{\stconn}{\textsc{stConn}}

\nc{\pstconn}{\ensuremath{p\textsc{-stConn}}}

\nc{\parity}{\textsc{Parity}}

\nc{\C}{\mathsf C}
\nc{\D}{\mathsf D}
\rnc{\PC}{\mathsf{PC}}

\nc{\FO}{\textup{FO}}

\nc{\phalt}{\ensuremath{p\textsc{-Halt}}}

\nc{\NTM}{\textup{NTM}}

\nc{\cn}{\ensuremath{\omega}}

\nc{\DTdepth}{\textup{DTdepth}}
\nc{\DTdepthv}{\ensuremath{\DTdepth_{\rm vertex}}}

\nc{\pgapclique}[1]{\ensuremath{p\textsc{-Gap}_{#1}\textsc{-Clique}}}
\nc{\pgapwsat}[1]{\ensuremath{p\textsc{-Gap}_{#1}\textsc{-Wsat}}}

\rnc{\log}{\textup{log}}

\nc{\res}[1]{\ensuremath{\!\!\upharpoonright_{#1}}}

\nc{\dlogtime}{\textup{dlogtime}}

\nc{\pacr}{\textrm{para-$\AC^0$}}
\nc{\pac}{\textrm{pac}}
\nc{\pwac}{\textrm{pwac}}

\nc{\LFP}{\textup{LFP}}

\nc{\G}{\mathbf G}

\begin{document}

\title{Some lower bounds in parameterized $\AC^0$}
\author{Yijia Chen\\\normalsize School of Computer Science\\
\normalsize Fudan University\\
\normalsize yijiachen@fudan.edu.cn\\
\and
J\"{o}rg Flum\\\normalsize Mathematisches Institut \\
\normalsize Universit\"{a}t Freiburg\\
\normalsize joerg.flum@math.uni-freiburg.de}

\date{}
\maketitle

\begin{abstract}
We demonstrate some lower bounds for parameterized problems via
parameterized classes corresponding to the classical $\AC^0$. Among
others, we derive such a lower bound for all fpt-approximations of the
parameterized clique problem and for a parameterized halting problem,
which recently turned out to link problems of computational complexity,
descriptive complexity, and proof theory. To show the first lower bound,
we prove a strong $\AC^0$ version of the planted clique conjecture:
$\AC^0$-circuits asymptotically almost surely can not distinguish between
a random graph and this graph with a randomly planted clique of any size
$\le n^\xi$ (where $0 \le \xi < 1$).
\end{abstract}

\section{Introduction}
For $k\in \mathbb N$ the $k$-clique problem asks, given a graph $G$, whether
it contains a clique of size $k$. In~\cite{ros08}, Rossman showed that the
$k$-clique problem has no bounded-depth and unbounded-fan-in circuits of
size $O(n^{k/4})$, where $n$ is the number of vertices in an input graph.
Therefore, there doesn't exist a family $\left(\C_{\binom{n}{2},
k}\right)_{n,k\in \mathbb N}$ of circuits such that for some functions $d,f:\mathbb N\to \mathbb N$,
\begin{itemize}
\item every $\C_{\binom{n}{2},k}$ has depth at most $d(k)$ and size
    bounded by $f(k)\cdot n^{k/4}$,

\item an $n$-vertex graph $G$ has a $k$-clique if and only if
    $\C_{\binom{n}{2},k}(G)=1$. Here $\C_{\binom{n}{2}}$ has an input node
    for every potential edge.
\end{itemize}
If the constraint on the depth of the circuits could be removed, then we
would immediately obtain that the \emph{parameterized clique problem}
\npprob[7]{\pclique}{A graph $G$ and $k\in \mathbb N$}{$k$}{Does $G$
contain a clique of size $k$?}
cannot be solved in time $f(k)\cdot n^{O(1)}$. Thus, \pclique\ would not be
fixed-parameter tractable (\FPT) and hence,  $\FPT\ne \W 1$, since
$\pclique$ is in the parameterized class \W 1. Therefore, Rossman's result
may be viewed as  an $\AC^0$ version of $\FPT\ne \W 1$, an inequality
conjectured by most experts of the field (recall that the complexity class
$\AC^0$ contains all problems that can be computed by bounded-depth and
unbounded fan-in circuits of polynomial size).

In \cite{elbstotan15} Elberfeld et al.\ introduced the parameterized class
\paraAC\ as the $\AC^0$ analog of the class \FPT: A problem is in \paraAC\
if it can be computed by \emph{dlogtime-uniform} $\AC^0$-circuits
after an (arbitrarily complex) \emph{precomputation}~\cite{flugro03} on the
parameter. Later in~\cite{banstotan15} it was shown that \paraAC\ contains
the \emph{parameterized vertex cover problem} (\pvc), one of the
archetypal fixed-parameter tractable problems.
%
%
For various other problems the authors of \cite{banstotan15} also proved
their membership in \paraAC. Concerning nonmembership, a result
in~\cite{beaimppit98} shows that the parameterized $st$-connectivity problem
(\pstconn), i.e., the problem of deciding whether there is a path of length
at most $k$ between vertices $s$ and $t$ in a graph $G$, parameterized by
$k$, is not in \paraAC. It is worth noting that $st$-connectivity is
solvable in polynomial time, and hence $\pstconn\in \FPT$.

The class $\AC^0$ is one of the best understood classical complexity
classes. Already in~\cite{ajtai83, fursaxsip84} it was shown that  \parity,
the problem  of deciding whether a binary string contains an  even number of
$1$'s, is not in $\AC^0$. Since \parity\ has a very low complexity, for many
other problems, including \vc\ and \clique,  the $\AC^0$-lower bound  can be
easily derived by reductions from \parity. Similarly, as $\pclique\notin
\paraAC$, it is not very hard to see, using some appropriate weak
parameterized reductions, that many other parameterized problems, including
the dominating set problem, are not in \paraAC.

It is well known that the class $\AC^0$ is intimately connected to
first-order logic (\FO). In fact, the problems decidable by
dlogtime-uniform  $\AC^0$-circuits are precisely those definable in
$\FO(<,+, \times)$, that is, in first-order logic for ordered structures
with built-in predicates of addition and multiplication.

Now we can also study various parameterized classes based on fragments of
$\FO(<,+, \times)$. Let us emphasize that this is not merely an academic
exercise. Logic and parameterized complexity are surprisingly intertwined
with each other, which, among others, is witnessed by various algorithmic
meta-theorems (see e.g.~\cite{grokresie14}). Moreover, the problem whether
there is a logic for \PTIME, a central problem of descriptive complexity,
turned out (see~\cite{cheflu12} for a thorough discussion)  to be related to
the complexity of the parameterized halting problem
\npprob[10.5]{\phalt}{$n\in \mathbb N$ in \emph{unary} and a
nondeterministic Turing machine (\NTM) $\mathbb M$}{$|\mathbb M|$, the size
of he machine $\mathbb M$}{Does $\mathbb M$ accept the empty input tape in
at most $n$ steps?}
In fact, already in~\cite{nasremvic05} it was shown that PTIME has a logic
if \phalt\ has an algorithm with running time $n^{f(|\mathbb M|)}$ for some
function $f$. We get a family $(\C_{n,k})_{n,k\in \mathbb N}$ of circuits
such that
\begin{itemize}
\item every $\C_{n,k}$ has depth 2 and size $g(k)\cdot n$ for some
    function $g:\mathbb N\to \mathbb N$,

\item an \NTM\ $\mathbb M$ accepts the empty input tape in at most $n$
    steps if and only if $\C_{n,|\mathbb M|}(n,\mathbb M)=1$
\end{itemize}
by hard-wiring into $\C_{n,k}$ the \NTM s of size $k$ which halt on empty
input in $\le n$ steps.

Therefore, $\phalt$ is in a \emph{nonuniform} version of \paraAC. So the
question arises whether $\phalt\in \paraAC$. Note that a positive answer
will yield that $\phalt\in \FPT$, which is considered to be highly
unlikely~\cite{cheflu12}. Hence, the goal is to show \emph{unconditionally}
that $\phalt\notin \paraAC$. To the best of our knowledge, all existing
$\AC^0$ lower bounds apply to both uniform and nonuniform circuits. Perhaps,
in order  to settle the complexity of $\phalt$ with respect to \paraAC,  a
better understanding of the uniformity conditions of circuits is really
required.

\subsection*{Our work}
In this paper, we systematically investigate lower bounds in terms of
\paraAC. We show that a number of problems are not in this class or in some
of its proper subclasses. To some extent, our results appear rather
separated and our proofs are often built on known results and techniques.
Nevertheless, as \emph{unconditional lower bounds} are still rare in
parameterized complexity, \paraAC\ is in our opinion the best starting point
for this line of research.

Following the framework proposed in~\cite{flugro03}, we first compare two
possible definitions of \paraAC\ depending on different ways to obtain
parameterized classes from classical ones. We have already mentioned the
first one, in which an arbitrary precomputation can be performed on the
parameter before a standard computation according to the corresponding
classical class.  The second approach requires  the parameterized problem to
be in the classical class if we restrict to instances were  the parameter is
far smaller than the size of the input. We show that both views lead to the
same \paraAC.

Then we derive a first set  of lower bound results: We show that many
natural \W 1-hard problems are not in \paraAC\ by arguing that the
corresponding reductions from \pclique\ can be made in~$\AC^0$. Among
others, they include the  weighted satisfiability problems for classes of
propositional formulas, which define the W-hierarchy.

We present a modeltheoretic tool, based on the color-coding method, which
allows to show membership in $\AC^0$ (similarly as done
in~\cite{banstotan15} via circuits).

We generalize Rossman's result mentioned at the beginning of this
introduction and show that any \fpt-approximation of \pclique\ is not in
\paraAC. To get this result we prove that $\AC^0$-circuits asymptotically
almost surely can not distinguish between a random graph and this graph with
a randomly planted clique of any size $\le n^\xi$ with $0 \le \xi < 1$. Our
first proof of the last two results used the sophisticated machinery
in~\cite{ros08}. Here we outline a proof, suggested to us anonymously, which
is directly built on Beame's \emph{Clique Switching Lemma}~\cite{bea94}. The
\fpt-approximation lower bound of \pclique\ again can be transferred to the
weighted satisfiability problems, provided the propositional formulas are of
odd depth.


Finally we turn to \phalt. We are not able to show $\phalt\notin \paraAC$,
however, using the decidability of Presburger's arithmetic we prove that
\phalt\ is not in para-$\FO(<,+)$, not even in $\X\FO(<,+)$. On the other
hand, $\phalt\in\textup{nonuniform-}\para\FO(<,+)$.

\section{Preliminaries}\label{sec:pre}

By $\mathbb N$ we denote the set of nonnegative integers. For every $n\in
\mathbb N$ we let $[n]:= \{1, \ldots, n\}$. Moreover, let~$\mathbb R$ be the
set of real numbers, $\mathbb R_{+}:= \big\{r\in \mathbb R\bigmid
r>0\big\}$, and $\mathbb R_{\ge 1}:= \big\{r\in \mathbb R\bigmid r\ge
1\big\}$. For any set~$A$ and $k\in \mathbb N$ we define $\binom{A}{k}$ as
the class of $k$-element subsets of $A$, i.e., $\big\{S\subseteq A\bigmid
|S|= k\big\}$.

\medskip
A (simple) graph $G= (V(G),E(G))$ \big(for short, $G=(V,E)$\big) is
undirected and has no loops and multiple edges. Here, $V(G)$ is the vertex
set and $E(G)$ the edge set, respectively. A subset $C\subseteq V(G)$ is a
\emph{clique} of $G$ if for every $u,v\in C$ either $u=v$ or $\{u,v\}\in
E(G)$. And $D\subseteq V(G)$ is a \emph{dominating set} of $G$ if for every
$v\in V(G)$ either $v\in D$ or there exists $u\in D$ with $\{u,v\}\in E(G)$.

\subsection*{Relational structures and first-order logic}
A \emph{vocabulary} $\tau$ is a finite set of relation symbols. Each
relation symbol has an \emph{arity}. A \emph{structure}~$\str{A}$ of
vocabulary $\tau$, or simply structure, consists of a finite set $A$ called
the \emph{universe}, and an interpretation $R^{\str{A}}\subseteq A^r$ of
each $r$-ary relation symbol $R \in \tau$. For example, a graph $G$ can be
identified with a structure $\mathcal A(G)$ of vocabulary $\{E\}$ with
binary relation symbol $E$ such that $A(G):= V(G)$ and $E^{\mathcal A(G)}:=
\{(u,v) \mid \{u,v\}\in E(G)\}$.

Formulas of first-order logic of vocabulary $\tau$ are built up from atomic
formulas $x=y$ and $Rx_1 \ldots x_r$, where $x,y,x_1,\ldots, x_r$ are
variables and $R\in\tau$ is of arity $r$, using the boolean connectives and
existential and universal quantification. For example, for every $k\ge 1$
let
\[
\textit{clique}_k := \exists x_1 \ldots \exists x_k
 \Big(\bigwedge_{1\le i<j \le k} (\neg x_i= x_j \wedge Ex_ix_j)\Big).
\]
Then a graph $G$ has a $k$-clique if and only if $\str A(G) \models
\textit{clique}_k$.



\subsection*{Parameterized complexity}
We fix an alphabet $\Sigma:= \{0,1\}$. A \emph{parameterized problem} $(Q,
\kappa)$ consists of a classical problem $Q\subseteq \Sigma^*$ and a
function $\kappa: \Sigma^*\to \mathbb N$, the \emph{parameterization},
computable in polynomial time. As an example, we have already seen
$\pclique$ in the Introduction. A similar problem is the \emph{parameterized
dominating set problem}.
\npprob[7]{$\pds$}{A graph $G$ and $k\in \mathbb N$}{$k$}{Does $G$ contain a
dominating set of size $k$?}
Both, $\pclique$ and $\pds$, play an important role in parameterized
complexity, mainly because they are complete for the classes $\W 1$ and $\W
2$, respectively. Recall that the classes of the W-hierarchy are defined by
taking the closure under \fpt-reductions of the following weighted
satisfiability problem for suitable classes $\Gamma$ of propositional
formulas or circuits.
\npprob[9.5]{$\pwsat(\Gamma)$}{$\gamma\in \Gamma$ and $k\in \mathbb
N$}{$k$}{Does $\gamma$ have a satisfying assignment of Hamming weight $k$?}

\begin{defn}\label{def:fptred}
Let $(Q, \kappa)$ and $(Q', \kappa')$ be two parameterized problems. An
\emph{fpt-reduction} from $(Q, \kappa)$ to $(Q', \kappa')$ is a mapping $R:
\Sigma^* \to \Sigma^*$ such that:
\begin{enumerate}
\item[--] For all $x \in \Sigma^*$ we have $\big (x \in Q \iff R(x) \in Q'
    \big)$.

\item[--] For all $x \in \Sigma^*$, the image $R(x)$ is computable in time
    \[
    f(\kappa(x)) \cdot |x|^{O(1)}
    \]
    for a computable $f: \mathbb N\to \mathbb N$.

\item[--] There is a computable function $g : \mathbb N\to \mathbb N$ such
    that $\kappa'(R(x)) \le g(\kappa(x))$ for all $x \in \Sigma^*$.
\end{enumerate}
If there is an \fpt-reduction from $(Q,\kappa)$ to $(Q', \kappa')$, then we
write $(Q,\kappa)\le^{\fpt} (Q', \kappa')$.
\end{defn}

For $t \ge 0$ and $d \ge 1$ we inductively define the following classes
$\Gamma_{t,d}$ and $\Delta_{t,d}$ of propositional
formulas:
\begin{align*}
\Gamma_{0,d} & := \ \{\lambda_1 \wedge \ldots \wedge \lambda_c \mid c \le d,
\lambda_1, \ldots, \lambda_c \ \text{literals} \}, \\
\Delta_{0,d} & := \ \{\lambda_1 \vee \ldots \vee \lambda_c \mid c \le d,
\lambda_1, \ldots, \lambda_c \ \text{literals}\}, \\
\Gamma_{t+1,d} & := \ \Big \{\bigwedge_{i \in I} \delta_i \mid \text{$I$
finite, $\delta_i \in \Delta_{t,d}$ for all $i \in I$} \Big \}, \\
\Delta_{t+1,d} & := \ \Big \{\bigvee_{i \in I} \gamma_i \mid \text{$I$ finite,
$\gamma_i \in \Gamma_{t,d}$ for all $i \in I$} \Big \}.
\end{align*}
Now we are ready to  define the classes of the \emph{W-hierarchy}.
\begin{defn}\label{def:Wclass}
Let $t \ge 1$. Then
\[
\W t := \bigcup_{d \ge 1} \big \{(Q, \kappa)
\mid (Q, \kappa) \le^{\fpt} \pwsat(\Gamma_{t,d}) \big \}.
\]
\end{defn}

\subsection*{Circuit Complexity}
A circuit $\C$ with $n$ input gates is a directed acyclic graph in which
every node (i.e., gate) is labelled by $\bigwedge$, $\bigvee$, $\neg$, or by
one of the variables, or by $0$ or $1$. All $\bigwedge$ and $\bigvee$ gates may
have arbitrarily many inputs, i.e., $\C$ is of \emph{unbounded fan-in}. The
\emph{depth} of $\C$ is the length of a longest directed path in~$\C$. The
\emph{size} of $\C$, denoted by $|\C|$, is the number of gates in $\C$. We
often tacitly identify $\C$ with the function $\C: \{0,1\}^n\to \{0,1\}^m$
it computes. Here, $n$ is the number of variables of $\C$ and $m$ the number
of its \emph{output gates}.

$\AC^0$ is the class of problems that can be computed by circuits of
bounded-depth and polynomial size. More precisely:

\begin{defn}
Let $Q\subseteq \Sigma^*$. We say that $Q\in \AC^0$ if there exists a family
of boolean circuits $(\C_n)_{n\in \mathbb N}$ such that:
\begin{description}
\item[(A1)] The depth of every $\C_n$ is bounded by a fixed constant.

\item[(A2)] $|\C_n|= n^{O(1)}$.

\item[(A3)] Let $x\in \Sigma^*$. Then ($x\in Q$ if and only if
    $C_{|x|}(x)= 1$). In particular,  $\C_n$ has $n$ input gates.

\item[(A4)] $(\C_n)_{n\in \mathbb N}$ is \emph{dlogtime-uniform}, that is:
    there is a \emph{deterministic logtime Turing machine}~$\mathbb M$
    which on input $1^n$ outputs the circuit~$\C_n$. More precisely,
    $\mathbb M$ recognizes the language
    \[
    \big\{(b,i,1^n) \bigmid \text{the
    $i$th bit of the binary encoding of $C_n$ is $b$}\big\}
    \]
    \big(cf. Section~6 of~\cite{barimm90}\big).
\end{description}
Often, $(\C_n)_{n\in \mathbb N}$ are called \emph{$\AC^0$-circuits}.
\end{defn}

We remark that most lower bounds in our paper still hold without the
requirement (A4). Therefore, (A4) is irrelevant for most of our results.
However, with this uniformity condition, $\AC^0$ characterizes precisely the
class of problems that are definable in $\FO(<, +, \times)$~\cite{barimm90}.

\section{\paraAC\ and Some Natural Examples}

\begin{defn}[\cite{banstotan15}]\label{def:paraAC}
Let $(Q, \kappa)$ be a parameterized problem. Then $(Q, \kappa)$ is in
\paraAC\ if there exists a family $\big(\C_{n,k}\big)_{n,k\in
\mathbb N}$ circuits such that:
\begin{enumerate}
\item[(P1)] The depth of every $\C_{n,k}$ is bounded by a fixed constant.

\item[(P2)] $|\C_{n,k}|\le f(k)\cdot n^{O(1)}$ for every $n,k\in \mathbb
    N$, where $f:\mathbb N\to \mathbb N$ is a computable function.

\item[(P3)] Let $x\in \Sigma^*$. Then \big($x\in Q$ if and only if
    $\C_{|x|, \kappa(x)}(x)=1$\big).

\item[(P4)] There is a deterministic Turing machine that on input $(1^n,
    1^k)$ computes the circuit $\C_{n,k}$ in time $g(k)+ O(\log\; n)$,
    where $g:\mathbb N\to \mathbb N$ is a computable function.
\end{enumerate}
\end{defn}

For future reference, we restate a \paraAC\ version of Rossman's main
result~\cite{ros08} as follows.
\begin{theo}\label{thm:ross}
Let $k\in \mathbb N$. Then there is no family $\left(\C_{\binom{n}{2}}\right)_{n\in \mathbb
N}$ of circuits such that the following conditions are all satisfied.
\begin{enumerate}
\item[--] The depth of every $\C_{\binom{n}{2}}$ is bounded by a fixed constant $d\in \mathbb N$.

\item[--] The size of $\C_{\binom{n}{2}}$ is $O(n^{k/4})$.

\item[--] Let $G$ be a graph and $n:= |V(G)|$. Then $G$ has a $k$-clique
    if and only if $\C_{\binom{n}{2}}(G)=1$. Here, $\C_{\binom{n}{2}}$ has
    an input node for every potential edge.
\end{enumerate}
In particular, $\pclique\notin \paraAC$.
\end{theo}

\begin{rem}
Recall that Chen et al.~\cite{chehua04} showed that $\pclique$ has no
algorithms of running time $f(k)\cdot |n|^{o(k)}$ unless the Exponential
Time Hypothesis (\ETH) fails. Theorem~\ref{thm:ross} in fact establishes an $\AC^0$ version of this result without using ETH.
\end{rem}

Next, we give two equivalent characterizations of $\paraAC$. The first one
\big(i.e., between (i) and (ii)\big) was already mentioned
in~\cite{elbstotan15}. Note that in~\cite{elbstotan15} it is required that a
problem in $\paraAC$ has an $\AC^0$ computable parameterization.

\begin{prop}\label{prop:equiv}
Let $(Q,\kappa)$ be a parameterized problem. Consider the following
statements.
\begin{enumerate}
\item[(i)] $(Q,\kappa)\in \paraAC$.

\item[(ii)] There is a \emph{precomputation}, that is, a computable
    function $\textit{pre}: \mathbb N\to \Sigma^*$ and $\AC^0$-circuits $\big(\C_n\big)_{n\in \mathbb N}$ such that for every $x\in \Sigma^*$,
    \begin{eqnarray*}
    x\in Q & \iff & \C_{|(x,\textit{pre}(\kappa(x))|}(x, \textit{pre}(\kappa(x)))= 1.
    \end{eqnarray*}

\item[(iii)] $Q$ is decidable, and there is a computable function
    $h:\mathbb N\to \mathbb N$ and $\AC^0$-circuits $\big(\C_n\big)_{n\in
    \mathbb N}$ such that for every $x\in \Sigma^*$ with $|x|\ge
    h(\kappa(x))$,
    \begin{eqnarray*}
    x\in Q & \iff & \C_{|x|}(x)= 1.
    \end{eqnarray*}
\end{enumerate}
Then (iii) $\Rightarrow$ (i) and (1) $\Leftrightarrow$ (ii). If, in addition, the
parameterization $\kappa$ can be computed by $\AC^0$-circuits, then (i)
$\Rightarrow$ (iii), i.e., all three statements are equivalent.
\end{prop}

\proof (i) $\Rightarrow$ (ii) \ Let $(Q, \kappa)\in \paraAC$ be witnessed by
a family $\big(\C_{n,k}\big)_{n,k\in \mathbb N}$ of circuits. Moreover, let
$f,g:\mathbb N\to \mathbb N$ be the corresponding computable functions in
(P2) and (P4). Without loss of generality, we assume that $g$ is increasing
and $f(k)= 2^{g(k)}$.

Then, we define the precomputation as $\textit{pre}(k):= (k, f(k))$. We need
to construct a family of circuits $\big(\D_m\big)_{m\in \mathbb N}$ such
that for every $x\in \Sigma^*$, $y:= \big(x,\textit{pre}(\kappa(x))\big)$, and $m:= |y|$
\begin{eqnarray}\label{eq:D}
x\in Q & \iff & \D_{m}(y)=1.
\end{eqnarray}
The circuit $\D_m$ is basically an $\bigvee$-gate on all possible
$\C_{n,k}$'s with $n+f(k)\le m$. On input
$y=\big(x,\textit{pre}(\kappa(x))\big)= \big(x, (\kappa(x),
f(\kappa(x)))\big)$,
it detects the input $x$ and the
parameter $k= \kappa(x)$, and then uses $\C_{n,k}$ to evaluate on~$x$.
Clearly~\eqref{eq:D} holds. Note the size of $\D_m$ can be bounded as
\[
|\D_m|\le O\left(\sum_{n+f(k)\le m} |\C_{n,k}|\right)
 = O\left(\sum_{n+f(k)\le m} f(k)\cdot n^{O(1)}\right)
 \le m^{O(1)},
\]
where the last equality is by (P2) and $f(k)\le m$. The \dlogtime-uniformity
of $\D_m$ is also easy to see by (P4) and $f(k)= 2^{g(k)}$.
%

\medskip
\noindent (ii) $\Rightarrow$ (i) \  Given $\AC^0$-circuits
$\big(\C_m\big)_{m\in \mathbb N}$ and a precomputation $\textit{pre}:
\mathbb N\to \Sigma^*$ as in (2), it is our goal to construct a family
$\big(\D_{m,k}\big)_{m,k\in \mathbb N}$ of circuits which satisfies (P1) --
(P4) in Definition~\ref{def:paraAC}. For every $m,k\in \mathbb N$ let
$\D_{m,k}$ simulate the circuit $\C_{m+|\textit{pre}(k)|}(\_,
\textit{pre}(k))$, i,e, we fix the second part of the input of
$\C_{m+|\textit{pre}(k)|}$ as $\textit{pre}(k)$. Then for every $x\in
\Sigma^m$
\[
\D_{m, \kappa(x)}(x)= 1
\iff \C_{m+|\text{pre}(\kappa(x))|}(x, \textit{pre}(k))= 1 \iff x\in Q.
\]
This establishes (P3). The conditions on the depth, the size, and the
uniformity of $\D_{m,k}$ are routine.

\medskip
\noindent (iii) $\Rightarrow$ (i) \ Let the $\AC^0$-circuits
$(\C_n)_{n\in\mathbb N}$ be as in (iii) and let $n, k\in \mathbb N$. By
assumption, if $n\ge h(k)$, then the circuit $\C_n$ satisfies that
\big($x\in Q \iff \C_n(x)=1$\big) for every $x\in \Sigma^n$ with $\kappa(x)=
k$. So we can take $\D_{n,k}:= \C_n$. Otherwise, $n< h(k)$, then we define
\[
\D_{n,k}(x) := \bigvee_{y\in Q\cap \Sigma^n} x=y,
\]
Here $x=y$ is the abbreviation of the circuit $\bigwedge_{i\in [n]} x_i=
y_i$, where every $x_i$ ($y_i$) is the $i$th bit of $x$ ($y$, respectively).

\medskip
Now assume that there are $\AC^0$-circuits $\big(\PC_n\big)_{n\in \mathbb
N}$ such that for every $x\in \Sigma^*$ we have $\PC_{|x|}(x)= \kappa(x)$.
We show the direction from (i) to (iii). Let $\big(\C_{n,k}\big)_{n,k\in
\mathbb N}$, $f,g:\mathbb N\to \mathbb N$ be as stated in
Definition~\ref{def:paraAC} for (i). Again, we assume that $g$ is increasing
and $f(k)= 2^{g(k)}$. Now for every $n\in \mathbb N$ and $x\in\Sigma^n$ we
define
\[
\D_n(x)= \bigvee_{\substack{k\in \mathbb N\ \text{with}\\ f(k)\le n}}
 \big((\PC_n(x)= k)\wedge \C_{n,k}(x)\big).
\]
Then for every $x\in \Sigma^n$ with $k:= \kappa(x)$ and $|x|\ge f(k)$ it
holds
\[
x\in Q  \iff \C_{n,k}(x)=1 \iff \D_n(x)=1.
\]
It is easy to verify that $\big(\D_n\big)_{n\in \mathbb N}$ are
$\AC^0$-circuits. \proofend

In order to use Theorem~\ref{thm:ross} to show \paraAC\ lower bounds for
other problems, we introduce a more restricted form of fpt-reductions.
\begin{defn}\label{def:pacred}
Let $(Q, \kappa)$ and $(Q', \kappa')$ be two parameterized problems. A
\emph{\pacr-reduction} from $(Q, \kappa)$ to $(Q', \kappa')$ is a mapping
$R: \Sigma^* \to \Sigma^*$ such that:
\begin{enumerate}
\item[(R1)] For all $x \in \Sigma^*$ we have $\big(x \in Q \iff R(x) \in
    Q' \big)$.

\item[(R2)] There is a family of circuits $\big(\C_{n,k}\big)_{n,k\in
    \mathbb N}$, whose depth is bounded by a fixed constant, such that
    \begin{enumerate}
    \item for all $x \in \Sigma^*$, $\C_{|x|,\kappa(x)}(x)$ outputs
        $R(x)$;

    \item $\big|\C_{n,k}\big|\le f(k)\cdot |x|^{O(1)}$ for a computable
        function $f:\mathbb N\to \mathbb N$;

    \item there is a deterministic Turing machine that on input $(1^n,
        1^k)$ computes the circuit $\C_{n,k}$ in time $g(k)+ O(\log\;
        n)$, where $g:\mathbb N\to \mathbb N$ is a computable function.

    \end{enumerate}

\item[(R3)] There is a computable function $h: \mathbb N\to \mathbb N$
    such that $\kappa'(R(x)) \le h(\kappa(x))$ for all $x \in \Sigma^*$.
\end{enumerate}
If there is a \pacr-reduction from $(Q,\kappa)$ to $(Q', \kappa')$, then we
write $(Q,\kappa)\le^{\pac} (Q', \kappa')$.
\end{defn}

\medskip
However, in general $\paraAC$ is \emph{not} closed under $\pacr$-reductions
as witnessed by the following example.
\begin{exa}
Define
\[
Q:= \left\{(x,b) \left| \text{$x\in \{0,1\}^*$ and $b= \sum_{i\in [|x|]} x_i \mod 2$}
 \right .\right\}.
\]
Clearly, $Q$ is equivalent to the classical \parity\ problem of deciding
whether there is an even number of $1$'s in $x$. Thus $Q\notin \AC^0$.

We define two polynomial time computable parameterizations of $Q$ by
\begin{eqnarray*}
\kappa_1(x,b):= 0
 & \text{and} &
\kappa_2(x,b):= \sum_{i\in [|x|]} x_i \mod 2.
\end{eqnarray*}
Then it is easy to see that $(Q, \kappa_1)\notin \paraAC$ and $(Q,
\kappa_2)\in \paraAC$; yet $(Q, \kappa_1)\le^{\pac} (Q, \kappa_2)$ by the
identity mapping $R(x,b)= (x,b)$.

Note $(Q, \kappa_2)$ also serves as a counterexample for the direction from
(i) to (iii) in Proposition~\ref{prop:equiv}.
\end{exa}

Therefore we need a further requirement on $\pac$-reductions. The previous
example suggests to require the $\AC^0$-computability of the
parameterization (as done in~\cite{elbstotan15}). In fact, $\paraAC$ is
closed under those reductions. However, we choose another requirement, which
is simpler to verify and is satisfied by almost all natural reductions.

\begin{defn}\label{def:pwacred}
Let $(Q, \kappa)$ and $(Q', \kappa')$ be two parameterized problems. A
\emph{weak \pacr-reduction} from $(Q, \kappa)$ to $(Q', \kappa')$ is a
\pacr-reduction which satisfies:
\begin{enumerate}
\item[(R3')] There is a computable function $h: \mathbb N\to \mathbb N$
    such that $\kappa'(R(x))= h(\kappa(x))$ for all $x \in \Sigma^*$.
\end{enumerate}
$(Q,\kappa)\le^{\pwac} (Q', \kappa')$ means that there is a weak
\pacr-reduction from $(Q,\kappa)$ to $(Q', \kappa')$.
\end{defn}

It is straightforward to verify that $\paraAC$ is closed under weak
$\pacr$-reductions.
\begin{lem}\label{lem:pacred}
Let $(Q, \kappa)$ and $(Q', \kappa')$ be two parameterized problems with
$(Q, \kappa)\le^{\pwac} (Q', \kappa')$. If $(Q', \kappa')\in \paraAC$, then
$(Q, \kappa)\in \paraAC$, too.
\end{lem}

%

\medskip
It is well known that $\pclique$ is \fpt-reducible to $\pds$. The reduction
presented in the next proof is even a weak $\pacr$-reduction and thus, by
Theorem~\ref{thm:ross} and the previous lemma yields:
\begin{prop}
$\pds\notin \paraAC$.
\end{prop}

\proof By the previous remark it suffices to present a weak \pacr-reduction
from $\pclique$ to $\pds$. Let $(G,k)$ be an instance of $\pclique$ with
$G=(V,E)$. We may assume that $E$ is not empty. Let $k\ge 2$. We construct a graph $H=(W,F)$ with
\begin{eqnarray}\label{eq:rcld}
\text{$G$ has a $k$-clique}
 & \iff &
\text{$H$ has a dominating set of size $k+{k\choose 2}$}.
\end{eqnarray}
Let $\textup{new}(i)$ and $\textup{new}(i,j)$ with $i,j\in [k]$ and $i<j$ be
new vertices. The vertex set $W$ of $H$ is the disjoint union of three types
of sets:
\begin{enumerate}
\item[(a)] $\{\textup{new}(i)\}\cup V(i)$ for $i\in[k]$, where each $
    V(i)$ is a (disjoint) copy of $V$;

\item[(b)] $\{i\}\times \{j\} \times V(i)\times V(j)$ for $i,j\in [k]$
    with $i<j$;

\item[(c)] $\{\textup{new}(i,j)\}\cup E(i,j)$ for $i,j\in [k]$ with $i<j$,
    where each $E(i,j)$ is a (disjoint) copy of the edge set~$E$.
\end{enumerate}
We denote by $v(i)$ the copy of $v\in V$ in $V(i)$ and by $e(i,j)$ the copy
of $e\in E$ in $E(i,j)$. The set $F$ consists of the following edges:
\begin{enumerate}
\item[(d)] for $i\in[k]$ edges that make $\{\textup{new}(i)\}\cup V(i)$ a
    clique;

\item[(e)] for $i,j\in [k]$ with $i<j$ edges that make
    $\{\textup{new}(i,j)\}\cup E(i,j)$ a clique;

\item[(f)] for $i,j\in [k]$ with $i<j$ and every $(i,j,u(i),v(j))\in \{i
    \}\times \{j\}\times V(i)\times V(j)$ an edge from this vertex to
    every $u'(i)\in V(i)$ with $u\ne u'$ and an edge to every $v'(j)\in
    V(j)$ with $v\ne v'$; furthermore, if $\{u,v\}\in E$, then an edge
    from $(i,j,u(i),v(j))$ to the vertex $\{u,v\}(i,j)$ (in $E(i,j)) $.
\end{enumerate}
Then the equivalence~\eqref{eq:rcld} holds. In fact, first assume that $C:=
\{u_1,\ldots, u_k\}$ is a $k$-clique in~$ G$. Then the set
\[
D(C):= \big\{u_1(1),\ldots, u_k(k)\big\}
 \cup \big\{\{u_i,v_j\}(i,j) \mid \text{$ i,j\in [k]$ and $i<j$}\big\}
\]
is a dominating set in~$H$.

Conversely, assume $D$ is a  dominating set in $H$ of size $k+{k\choose 2}$.
In view of the elements of the form $\textup{new}(i)$ and
$\textup{new}(i,j)$, we see that $D$ must contain elements of each
$\{\textup{new}(i)\}\cup V(i)$ and of each $\{i\}\times \{j\}\times
V(i)\times V(j) $. Thus, $D$ consists of exactly one element of each of
these sets. Note that the element from $\{\textup{new}(i)\}\cup V(i)$ must
be distinct from $\textup{new}(i)$, as otherwise at most one element from
every $\{i\}\times \{j\}\times V(i)\times V(j)$ can be dominated by $D$ (but
$|V|\ge 2$ as, by assumption, $E\ne\emptyset)$. So let $u_i(i)$ with $u_i\in
V$ be the element of $D$ in $V(i)$. As $D$ dominates the element
$(i,j,u_i(i),u_j(j))$, we see that it has to be dominated by some element of
$ E(i,j)$; that is, $\{u_i,u_j\}\in E$. Thus $\{u_1,\ldots, u_k\}$ is a
clique. \proofend

\begin{cor}\label{cor:cliquedswsat}
Let $t,d\ge 1$ with $t+d\ge 3$. Then $\pwsat(\Gamma_{t,d})\notin
\paraAC$.
\end{cor}

\proof For every graph $G= (V,E)$ we define a propositional formula
\[
\delta_G:= \bigwedge_{\substack{\text{$u,v\in V$ with}\\ \text{$u\ne v$ and $\{u,v\}\notin E$}}}
\neg X_u \vee \neg X_v.
\]
Clearly, for every $k\in \mathbb N$,
\begin{eqnarray}\label{eq:cliquered}
\text{$G$ has a $k$-clique} & \iff &
\text{$\delta_G$ has a satisfying assignment of weight $k$}.
\end{eqnarray}
This gives a weak \pacr-reduction from \pclique\ to $\pwsat(\Gamma_{1,2})$,
or $\pwsat(\Gamma_{t,1})$ in case $t\ge 2$. \proofend

Similarly, one can show that basic problems like $p\textsc{-Hom}$,
$p\textsc{-Emb}$, $p\textsc{-Subgraph-Isomorphism}$, and
$p\textsc{-MC}(\Sigma^1_1)$ are not in $\paraAC$ (we use the notations
of~\cite{flugro03}).

\medskip
In view of Corollary~\ref{cor:cliquedswsat} the reader might wonder about
the status of $\pwsat(\Gamma_{1,1})$. Using the color-coding technique as
in~\cite{banstotan15}, one can show that the problem is in fact solvable in
$\paraAC$. We present a more logic-oriented technique for such proofs. It
uses $\FO(<,+, \times)$ instead of \dlogtime-uniform $\AC^0$. First, we
recall the following lemma from \cite[page~349]{flugro06}:

\begin{lem}\label{lem:cc}
For every sufficiently large $n\in \mathbb N$, it holds that for all $k\le
n$ and for every $k$-element subset $X$ of $[n]$, there exists a prime $p <
k^2\cdot \log_2\; n$ and $q< p$ such that the function $h_{p,q}: [n]\to \{0,
\ldots, k^2 -1\}$ given by $h_{p,q}(m):= (q\cdot m \mod p) \mod k^2$ is
injective on $X$.
\end{lem}

For $n\in \mathbb N$ denote by  $<^{[n]}$ the natural ordering on $[n]$.
Clearly, if $\str{A}$ is any ordered structure, then $(A,<^{\str{ A}})$ is
isomorphic to $([|A|],<^{[|A|]})$ and the isomorphism is unique.
Furthermore, for ternary relation symbols $+$ and $\times$ we consider the
ternary relations $+^{[n]}$ and $\times^{[n]}$ on $[n]$ that are the
relations underlying the addition and the multiplication of $\mathbb N$
restricted to $[n]$. That is,
\begin{equation}\label{eq:pm}
\begin{array}{rl}
+^{[n]}& := \big\{(a,b,c)\bigmid \text{$a,b,c\in [n]$ with $c=a+b$}\big\}, \\[1mm]
\times^{[n]}& := \big\{(a,b,c)\bigmid \text{$a,b,c\in [n]$ with $c=a\cdot b$}\big\}.
\end{array}
\end{equation}
Let $\tau$ be a vocabulary which does not contain the relation symbols
$<,+,\times$ and set $\tau_{<,+, \times}:= \tau \cup \{<$, $+,\times\}$. We
say that a $\tau_{<,+, \times}$-structure $\str{A}$ \emph{has built-in
addition and built-in multiplication} if $(A, <^{\str{ A}}$, $+^{\str{ A}},
\times^{\str{A}})$ is isomorphic to $([|A|], <^{[|A|]}, +^{[|A|]},
\times^{[|A|]})$. Sometimes we write $\varphi\in \FO(<, +, \times)$ to
emphasize that $\varphi$ is a first-order formula in a vocabulary containing
the symbols $<,+,\times$.

\begin{cor}\label{cor:cc}
There is a computable function which associates every $k\in \mathbb N$ with
a structure $\str C(k)$ and every \FO-formula $\varphi(x)$ with an
$\FO(<,+,\times)$-sentence $\chi_{\varphi}$ such that for every structure
$\str A$,
\begin{equation}\label{eq:ccvarphi}
\begin{array}{l}
[\str{ A}:\ \str C(k)]  \models \chi_{\varphi}\\[1mm]
{\hspace{.5cm}}\iff \text{there are pairwise distinct $x_1, \ldots, x_k\in A$ with
 $\str A\models \varphi(x_i)$ for every $i\in [k]$}.
\end{array}
\end{equation}
Here, $[\str{ A}:\str C(k)]: =\str B= \big(\str A\dotcup \str C(k), U^{\str
B}, <^{\str B}, +^{\str B}, \times^{\str B}\big)$ is defined as follows.
\begin{itemize}
\item $\str A\dotcup \str C(k)$ is the disjoint union of $\str A$ and
    $\str C(k)$.

\item $U^{\str B}:= A$.

\item $<^{\str B}$ is an ordering of $B$ and every element of $A$ precedes
    all elements of $C(k)$. Furthermore $<^{\str B}$ extends the ordering
    $\prec^{\str C(k)}$ given in $\str C(k)$.

\item $\str{B}$ has built-in addition $+^{\str B}$ and multiplication $\times^{\str B}$ .
 \end{itemize}
\end{cor}

\proof Let $\tau_0:=\{K, \prec, F\}$ with unary $K$, binary $\prec$, and
ternary $F$. We first define $\str C= \str C(k)$, a $\tau_0$-structure,
which basically embodies all functions from $\big\{0,\ldots, k^2-1\big\}$ to
$\{0,\ldots, k-1\}$.
\begin{itemize}
\item $C= \big\{0, \ldots, k^{k^2}-1\big\}$.

\item $K^{\str C}:= \{k-1\}$ is the singleton set containing the $k$-th
    element in $\str C$.

\item Let $\prec^{\str C}$ is the natural ordering on $C$.

\item Let $g_0, \ldots, g_{k^{k^2}-1}$ be an enumeration of all functions
    from $\big\{0,\ldots, k^2-1\big\}$ to $\{0,\ldots, k-1\}$. Then we
    define a ternary relation
    \[
    F^{\str C}:= \big\{(i,x,y) \bigmid g_i(x)= y\big\}.
    \]
\end{itemize}
Now let $\str{A}$ be any structure in a vocabulary $\tau$. We may assume
that $\tau\cap\tau_0=\emptyset$ by renaming symbols in $\tau$ if necessary.
Then, by the disjoint union $\str A\dotcup \str C(k)$ of $\str A$ and $\str
C(k)$ we mean the structure
\[
\big(A\dotcup C(k), (R^{\str{A}})_{R\in \tau}, (S^{\str{ C}(k)})_{S\in \tau_0}\big),
\]
where $A\dotcup C(k)$ is the disjoint union of the sets $A$ and $C(k)$.

\medskip
\noindent We view the universe of $\str B:=\str A\dotcup \str C(k)$ as
\[
B= \left\{0, \ldots, |A|+ k^{k^2}-1\right\}.
\]
In order to make formulas more readable, we introduce some abbreviations. We
freely use terms as in $x+y+z= w$ \big(which is equivalent to $\exists u (+
x y u \wedge + u z w)$\big). Note that $k-1$ can be defined in $\str{B}$ as
the unique $x$ satisfying the formula
\[
\exists u \exists v
\big(\neg U u \wedge \forall z (z< u \to Uz) \wedge Kv \wedge u+x= v\big).
\]
Thus, the number $k$, i.e., the $(k+1)$th element in $\str B$, can also be
easily defined. Clearly, $x = (y\!\! \mod z)$ is an abbreviation for
\[
 \exists u \big(x= u\times z+y \wedge y < z\big).
\]
Moreover, $g_i(x)= y$ is a shorthand for the formula
\begin{align*}
\gamma(i,x,y):= \exists u
\big(\neg U u & \wedge \forall z (z< u \to Uz) \\
 & \wedge \exists i' \exists x' \exists y'
  (i'=u+i\wedge x'=u+x \wedge y'=u+y \wedge Fi'x'y')\big).
\end{align*}
Now let
\[
\chi_{\varphi}:= \exists p \exists q \exists i \; \psi_{\varphi}(p,q,i),
\]
where
\[
\psi_{\varphi}(p,q,i)
  := \forall j \big(j<k \to \exists u (Uu \wedge \varphi^U(u) \wedge \rho(p,q,i,u,j)\big)
\]
\big(here $\varphi^U(u)$ is obtained from $\varphi(u)$ by relativizing all
quantifiers to $U$\big), and
\[
\rho(p,q,i,u,j) := g_i\big((q\times (u \!\! \mod p) \!\!\mod p) \!\!\mod k^2)=j\big).\ \footnotemark
\]
\footnotetext{Let us emphasize that $\rho(p,q,i,u,j)$ does not depend on $k$
which is defined from the unary relation $K^{\str C}$. Hence neither does
$\chi_{\varphi}$.}
\noindent Note that $\rho(p,q,i,u,j)$ is equivalent to
\[
g_i\big(h_{p,q}(u)\big)= j,
\]
where $h_{p,q}$ is defined in Lemma~\ref{lem:cc}. We replaced $(q\times u
\!\!\mod p)$ by $(q\times (u \!\! \mod p) \!\!\mod p)$, since $q\times u$
might exceed the size of $\str B$, i.e., $|A|+ k^{k^2}$.

We still need to show the equivalence~\eqref{eq:ccvarphi}. The direction
from right to left is easy, since $\str B\models \chi_{\varphi}$ means that
for some $p, q, i$ there exist $u_0, u_1, \ldots, u_{k-1} \in A$ with
\[
\str A\models \varphi(u_j) \quad \text{and} \quad g_i(h_{p,q}(u_j))=j
\]
for every $0\le j< k$. The second condition implies that all $u_j$'s are
distinct.

For the other direction, assume that there are $k$ elements $u_0, u_1,
\ldots, u_{k-1}\in A$ with $\str A\models \varphi(u_j)$ for all $j$. By
Lemma~\ref{lem:cc} there exist $p< k^2\cdot \log_2 n$ and $q< p$ such that
$h_{p,q}$ is injective on $\{u_0,\ldots ,u_{k-1}\}$. Since the range of
$h_{p,q}$ is $\{0,\ldots, k^2-1\}$, we can choose a function $g_i:
\{0,\ldots, k^2-1\}\to \{0,\ldots, k-1\}$ such that $g_i(h_{p,q}(u_j))= j$
for every $0\le j< k$. Since $q<p < k^2\cdot \log_2\; n$, we can guarantee
that
\[
(q\times (u \!\! \mod p)) < k^4 (\log_2 |A|)^2\le |A|+ k^{k^2}.
\]
Hence $\rho(p,q,u,j)$ gives the correct answer. \proofend

Let $\chi_{\varphi}^{-1}$ be the formula obtained by defining
$\psi_{\varphi}(p,q,i)$ by
\[
\psi_{\varphi}(p,q,i)
  := \forall j \big(j<(k-1) \to \exists u (Uu \wedge \varphi^U(u) \wedge \rho(p,q,u,j)\big),
\]
where the formula $\rho(p,q,u,j)$ remains unchanged. Then the last part of
the previous proof shows that
\begin{equation}\label{eq:ccvarphim}
\begin{array}{l}
[\str{ A}:\ \str C(k)]  \models \chi^{-1}_{\varphi}\\[1mm]
{\hspace{.2cm}}\iff \text{there are pairwise distinct $x_1, \ldots, x_{k-1}\in A$ with
 $\str A\models \varphi(x_i)$ for every $i\in [k-1]$}.
\end{array}
\end{equation}

\begin{prop}
$\pwsat(\Gamma_{1,1})\in \paraAC$
\end{prop}

\proof Let $\delta$ be a propositional formula in $\Gamma_{1,1}$. Clearly
$\delta$ has a satisfying assignment of Hamming weight $k$ if and only if in
$\delta$
\begin{enumerate}
\item[(i)] no propositional variable occurs both positively and
    negatively,

\item[(ii)] there are at least $k$ propositional variables,

\item[(iii)] there are \emph{no} $k+1$ variables which occur positively.
\end{enumerate}
Without loss of generality, we assume that (i) always holds. It is easy to
view $\delta$ as a structure $\str A(\delta)$ such that for some formulas
$\varphi_{\textup{var}}(x)$ and $\varphi_{\textup{pos}}(x)$ we have for
every propositional variable $Y$,
\begin{eqnarray*}
\str A(\delta)\models\varphi_{\textup{var}}(Z)
 & \iff & \text{$Z$ occurs in $\delta$}\\
\str A(\delta)\models\varphi_{\textup{pos}}(Z)
 &\iff & \text{$ Z$ occurs positively in $\delta$}.
\end{eqnarray*}
By Corollary~\ref{cor:cc} and \eqref{eq:ccvarphim} we see that
\[
\text{$\delta$ has a satisfying assignment of Hamming weight $k$}
 \iff [\str A(\delta):\str{ C(k+1)}]\models \big(\neg\chi_{\varphi_{\textup{pos}}}\wedge \chi^{-1}_{{\varphi_\textup{var}}}\big).
\]
This equivalence shows that the problem can be decided by $\FO(<,+, \times)$
after a precomputation on the parameter $k$. The result then follows from
Proposition~\ref{prop:equiv}. \proofend

\section{Inapproximability of $\pclique$ by \paraAC}

We recall the notion of \fpt\ approximation introduced in~\cite{chegro07}.
We present the definition for \pclique, the problem which interests us. It
can easily be generalized to any maximization problem.

\medskip
If not stated otherwise, $\rho:\mathbb N\to \mathbb R_{\ge 1}$ is always a
computable function such that the mapping $k\mapsto k/\rho(k)$ is
nondecreasing and unbounded.

\begin{defn}\label{def:paraapp}
An algorithm $\mathbb A$ is a \emph{parameterized approximation for
\pclique\ with approximation ratio $\rho$} if for every graph $G$ and $k\in
\mathbb N$ with $\cn(G)\ge k$ the algorithm $\mathbb A$ computes a clique
$C$ of $G$ such that
\[
|C|\ge \frac{k}{\rho(k)}.
\]
Here the clique number $\cn(G)$ is the size of a maximum clique of $G$. If
the running time of $\mathbb A$ is bounded by $f(k)\cdot |G|^{O(1)}$ where
$f:\mathbb N\to \mathbb N$ is computable, then $\mathbb A$ is an \emph{fpt
approximation algorithm}.
\end{defn}
We tend to believe that \pclique\ has no \fpt\ approximation algorithm for
any ratio $\rho$. Since \paraAC\ is a class of decision problems, in order
to prove a lower bound it is more convenient to deal with decision
algorithms instead of algorithms computing a clique.

\begin{defn}[\cite{chegro07}]\label{def:gap}
A decision algorithm $\mathbb A$ is a \emph{parameterized cost approximation
for \pclique\ with approximation ratio $\rho$} if for every graph $G$ and
$k\in \mathbb N$,
\begin{itemize}
\item if $k\le \cn(G)/ \rho(\cn(G))$, then $\mathbb A$ accepts $(G,k)$;

\item if $k> \cn(G)$, then $\mathbb A$ rejects $(G,k)$.
\end{itemize}
In other words, $\mathbb A$ decides the \emph{promise} problem:
\npprob{$\pgapclique \rho$}{A graph $G$ and $k\in \mathbb N$ such that
either $k\le \cn(G)/ \rho(\cn(G))$ or $k> \cn(G)$}{$k$}{Is $k\le \cn(G)/
\rho(\cn(G))$?}
\end{defn}
\noindent The intuition behind this definition: If $G$ contains a clique far
bigger than $k$, detecting a $k$-clique might become easier. It is
straightforward to verify that if \pclique\ has no parameterized \fpt\ cost approximation
of ratio $\rho$, then it has no parameterized \fpt\ approximation of ratio
$\rho$ either~\cite{chegro07}.

\medskip
\begin{theo}\label{thm:pgapcliqueparaAC}
Let $\rho:\mathbb N\to \mathbb R_{\ge 1}$ be a
computable function such that the mapping $k\mapsto k/\rho(k)$ is
nondecreasing and unbounded. Then
\[
\pgapclique \rho\notin \paraAC.
\]
\end{theo}
Our original proof of this result was based on a generalization of the
machinery developed in~\cite{ros08}, a generalization we first used to prove
that $\AC^0$ circuits are not sensitive to planted cliques of a reasonable
size, see Theorem~\ref{thm:indistAC}. The much simpler proof of
Theorem~\ref{thm:indistAC} we present here is based on Beame's Clique
Switching Lemma~\cite{bea94} and was suggested to us anonymously.

\subsection{Beame's Clique Switching Lemma}\label{sec:bea}
Let $n\in \mathbb N$. We consider graphs with vertex set $[n]$. To represent
functions on those graphs, every potential edge $e\in \binom{[n]}{2}$ is
encoded by a Boolean variable $X_e$. We set
\[
\mathcal X_n:= \left\{X_e \Bigmid e\in \binom{[n]}{2}\right\}.
\]
In particular, $X_e=1$ means that $e$ is present in the given graph,
otherwise $X_e=0$. Sometimes, it is convenient to understand $e$ as a
natural number with $e\in \left[\binom{n}{2}\right]$. Then, $e$ is the
$e$\/th potential edge in an $n$-vertex graph, and $X_e$ is the $e$th
variable in $\mathcal X_n$.

\medskip
For every $\ell\in [n]$ and $q\in\mathbb R$ with $0\le q\le 1$ let $\mu\in
\mathscr C^{\ell,q}_n$ be a \emph{random restriction}, $\mu:\mathcal X_n\to
\{0,1,\star\}$,
 generated as follows:
\begin{enumerate}
\item[--] Choose $U\in \binom{[n]}{\ell}$ uniformly at random
    and then set $\mu(X_e):= \star$ for every $e\in \binom{U}{2}$.

\item[--] For $e\notin \binom{U}{2}$ we set $\mu(X_e):= 1$ with
    probability $q$ and $\mu(X_e):=0$ with probability $1-q$.
\end{enumerate}
Let $F$ be a boolean function defined on the set of assignments from
$\mathcal X_n$ to $\{0,1\}$ and $\mu\in \mathscr C^{\ell,q}_n$. The function
$F\res{\mu}$ is defined on the set of assignments from $\mu^{-1}(\star)$ to
$\{0,1\}$ by:\
for any assignment $S: \mu^{-1}(\star)\to \{0,1\}$
\[
F\res{\mu}(S):= F(S\cup \mu),
\]
where $S\cup \mu: \mathcal X_n\to \{0,1\}$ is the assignment given by
\[
(S\cup \mu)(X_e):=
\begin{cases}
S(X_e) & \text{if $X_e\in  \mu^{-1}(\star)$} \\
\mu(X_e) & \text{otherwise}.
\end{cases}
\]
Recall that a rooted binary tree is a \emph{decision tree} on some variable
set $\mathcal X\subseteq \mathcal X_n$ if every leaf is labeled either $0$
or $1$, every internal node is labelled by a variable of $\mathcal X$, and
the edges between an internal node and its two children are labelled $0$ and
$1$. The \emph{vertex height} of a path $P$ in $T$ is the number of distinct
vertices occurring in edges $e$ such that {the corresponding $X_e$ appears
in $P$. The \emph{vertex height} $|T|_{v}$ of $T$ is the maximum vertex
height of a path in $T$.

For any boolean function $F$ as above, we set
\[
\DTdepthv(F)=
 \min \{|T|_{v} \mid \text{$T$ a decision tree computing $F$}\}.
\]

\begin{lem}[Beame's Clique Switching Lemma~\cite{bea94}\footnotemark]\label{lem:cliqueswitch}
Let $n,r\in \mathbb N$ and $0\le q\le 1/2$. Moreover, let $F$ be a
\DNF-formula of variable set $\mathcal X_n$ with conjunctive clauses of
vertex length at most $r$. For $s,\ell\in\mathbb N$ with $\ell:= pn$, where
$s\ge 0$ and $\ell:= pn$ with $p\le 1/(r (2/q)^{(r+s)/2})$, we have
\[
\Pr_{\mu\in \mathscr C^{\ell,q}_n}\Big[\DTdepthv \big(F\res{\mu}\big)> s\Big]
< \frac{8\big((2/q)^{(s+r-1)/2}pr\big)^s}{3}.
\]
Here, the \emph{vertex length} of a clause is the number of distinct
vertices in edges $e$ with $X_e$ appearing in this clause.
\end{lem}
\footnotetext{See the imbalanced version of~\cite[Lemma 3]{bea94} mentioned
in the first paragraph of page~12 of that paper.}

\medskip
We apply Lemma~\ref{lem:cliqueswitch} inductively on bounded-depth
circuits.
\begin{lem}\label{lem:pccAC}
Assume
\begin{itemize}
\item $k: \mathbb N\to \mathbb R_+$ with $k(n) \le \log_2\; n$ for all
    sufficiently large $n$ and $\lim_{n\to \infty} k(n)= \infty$,


\item $S: \mathbb N\to \mathbb N$ with $S(n)\ge n$,

\item $d: \mathbb N\to \mathbb N$.
\end{itemize}
Define $q:\mathbb N\to \mathbb R_+$ and $s:\mathbb N\to\mathbb N$ by
\begin{eqnarray}\label{eq:s}
q(n):= n^{-1/k(n)} & \text{and} &
 s(n):= \floor{\sqrt{k(n) (\log_n S(n) d(n))}},
\end{eqnarray}
and $\ell_i:\mathbb N\to\mathbb N$ inductively by
\begin{eqnarray}\label{eq:ellnn}
\ell_0(n) := n & \text{and} &
\ell_{i+1}(n) := \floor{\frac{\ell_i(n)}{n^{5 s(n)/k(n)}}}.
\end{eqnarray}
Then, $\ell_{d(n)}(n)= n^{1- \Theta\left(5d(n)\sqrt{(\log_n S(n)
d(n))/k(n)}\right)}$ and for every circuit $\C$ with variable set $\mathcal
X_n$, size bounded by $S(n)$, and depth bounded by $d(n)$, 
\[
\Pr_{\mu\in \mathscr C^{\ell_{d(n)}(n), q(n)}_n}
 \Big[\text{$\C\res{\mu}$ is constant} \Big] = 1- o(1).
\]
Moreover, the convergence rate can be bounded in terms of $S$, $d$, and $k$.
\end{lem}

\proof We fix an $n\in \mathbb N$ and let $k:= k(n)$, $q:= q(n)$, $S:=
S(n)$, $d:= d(n)$, $s:=s(n)$, and $\ell_i:=\ell_i(n)$. It is easy to see
that for every $i\in [d]$
\begin{equation}\label{eq:elli}
\frac{\ell_{i}}{\ell_{i+1}}\ge n^{5s/k}
\end{equation}
and
\begin{equation*}
\ell_d = n^{1- \Theta\left(5d\sqrt{(\log_n S d)/k}\right)}.
\end{equation*}
Let $\mu_0$ be the empty restriction, i.e., $\mu_0(X_e)= \star$ for every
$X_e\in \mathcal X_n$. For every $i\in [d]$ we let~$\pi_i$ be a random
restriction from $\mathscr C^{\ell_i, q}_{\ell_{i-1}}$. We set
\[
\mu_i:= \mu_{i-1} \circ \pi_i,
\]
where $\circ$ is defined in such a way that for every $X_e\in\mathcal X_n$,
\[
\mu_i(X_e)=
\begin{cases}
\mu_{i-1}(X_e) & \text{if $\mu_{i-1}(X_e) \ne \star$}, \\
\pi_i(X_{e'}) & \text{if $X_e$ is the $e'$th variable in $\mathcal X_{\ell_{i-1}}$
 with $\mu_{i-1}(X_e)= \star$ }.
\end{cases}
\]
It is easy to see that $\mu:= \mu_d$ has the distribution of $\mathscr
C^{\ell_d, q}_n$.

\medskip
\noindent Assume $n$ is sufficiently large. Hence~\eqref{eq:s} implies $s\ge
2$. Moreover, let
\begin{align*}
p_i:= \Pr \Big[\text{there is a gate $g$} & \text{ of depth $\le i$ with
               $\DTdepthv(\C_g\res{\mu_i})> s$} \\
  & \Bigmid \text{all gates of depth $\le i$ have $\DTdepthv(\C_g\res{\mu_{i-1}})\le s$}\Big],
\end{align*}
where $\C_g$ is the subcircuit of $\C$ with root $g$.

\medskip
Since every gate $g$ at height $0$ depends only on one edge variable $X_e$,
and $\mu_0(X_e)= \star$, we conclude that $\DTdepthv(\C_g\res{\mu_0})= 2$
and $p_0= 0$ (recall $s\ge 2$).

\medskip
Now assume that for all gates $g$ of depth $\le i$ we have
$\DTdepthv(C_g\res{\mu_i})\le s$. Let $g$ be an $\bigvee$-gate of depth $i+1
< d$. \footnote{In particular, $g$ is \emph{not} the output gate.} It
follows that $C_g\res{\mu_i}$ can be expressed as a \DNF-formula with terms
of vertex length at most $s$. By~\eqref{eq:elli}
\[
p:= \frac{\ell_{i+1}}{\ell_i}\le n^{-5s/k}.
\]
Then by Beame's Clique Switching Lemma (with $r\gets s$ and $s\gets s$) and
assuming $n$ is sufficiently large,
\begin{align*}
\Pr_{\pi_{i+1}\in \mathscr C^{\ell_{i+1}, q}_{\ell_i}}
 \Big[&\DTdepthv \big((C_g\res{\mu_i})\res{\pi_{i+1}}\big)> s\Big] \\
 & \le \frac{8}{3}\left((2/q)^{s-1/2} p s\right)^s \\
 & \le 3\left( s\big(2n^{1/k}\big)^{s-1/2} n^{-5s/k} \right)^s \\
 & \le 3n^{3 s^2/k} n^{-5 s^2/k} \\
 & = 3n^{-2 s^2/k} \\
 & \le 3n^{-1.9\cdot \log_n S d}
  \quad \qquad\text{\big(by~\eqref{eq:s} and for $n$, and hence $k$, sufficiently large\big)}\\
 & = \frac{3}{(S d)^{1.9}} \le \frac{o(1)}{Sd} \qquad \text{\big(by $S(n)\ge n$\big)}.
\end{align*}
%
To obtain the third inequality, i.e.,
\[
 3\left( s\big(2n^{1/k}\big)^{s-1/2} n^{-5s/k}\right)^s
 \le 3n^{3 s^2/k} n^{-5s^2/k},
\]
we argue, using $k\le \log_2\; n$,
\begin{align*}
s\big(2n^{1/k}\big)^{s-1/2}\le s\big(2n^{1/k}\big)^{s}
 = s 2^s n^{s/k} \le 2^{2 s} n^{s/k} \le n^{2 s/k} n^{3s/k}= n^{5 s/k}.
\end{align*}
Since there are at most $S$ many $\bigvee$-gates at depth $i+1$ in the
circuit $\C$, we have $p_{i+1}\le o(1)/d$ by a union bound.

\medskip
\noindent Finally, let $o$ be the output gate of $\C$ of depth $d$, i.e.,
$\C_o= \C$. Assume $\DTdepthv(\C_g\res{\mu_i}~)\le s$ for all gates $g$ of
depth $< d$. Again applying Beame's Clique Switching Lemma with parameters
$r\gets s$ and $s\gets 1$, we obtain
\begin{align*}
\Pr_{\pi_{d}\in \mathscr C^{\ell_{d}, q}_{\ell_{d-1}}}
 \Big[&\DTdepthv \big((\C_o\res{\mu_{d-1}})\res{\pi_{d}}\big)> 1\Big] \\
 & \le \frac{8}{3}\left((2/q)^{s/2} p s\right)
   < 3\left( s\big(2n^{1/k}\big)^{s/2} n^{-5 s/k}\right) \\
 & \le 3\left( 2^s\big(2n^{1/k}\big)^{s/2} n^{-5 s/k}\right)
  = 3\left( 2^{1.5 s} n^{s/2k} n^{-5 s/k}\right) \\
 & \le 3n^{2 s/k} n^{-5 s/k} = 3n^{-3 s/k}\le 3 \cdot 2^{-3 s} = o(1)
  \qquad \text{\big(by $\lim_{n\to \infty} k(n)= \infty$\big)}.
\end{align*}
Thus the probability of the event
\begin{quote}
either for a gate $g$ of depth $< d$ we have
$\DTdepth_v(C_g\upharpoonright_{\mu})> s$ or
$\DTdepth_v(C\upharpoonright_{\mu})> 1$
\end{quote}
is $o(1)$, which implies the desired result. \proofend

\subsection{A strong $\AC^0$ version of the planted clique conjecture}\label{sub:pla}
In the standard planted clique problem, we are given a graph $G$ whose edges
are generated by starting with a random graph with universe~$[n]$ and edge
probability $1/2$, then ``planting'' (adding edges to make) a random clique
on $k$ vertices; the problem asks for efficient algorithms finding such a
clique of size $k$. The problem was addressed
in~\cite{jer92,kuc95,alokrisu98}, among many others. It is conjectured that
no such algorithm exists  for $k= o(\sqrt{n})$. Here, as a consequence of Lemma~\ref{lem:pccAC},
we prove a statement considerably stronger than the $\AC^0$ version of this
conjecture.

\medskip
Let us be more precise. The \ERRE\ probability space $\ER(n,p)$, where $n\in
\mathbb N$ and $p\in\mathbb R$ with $0\le p\le 1$, is obtained as follows.
We start with the set $[n]$ of vertices. Then we choose every $e\in
\binom{[n]}{2}$ as an edge of $G$ with probability $p$, independently of the
choices of other edges.

For $G\in \ER(n,1/2)$ the expected size of a maximum clique is approximately
$2\, \log\; n$. Therefore~$G$ almost surely has no clique of size, say, $4\,
\log\; n$. For any graph $G$ with vertex set $[n]$ and any $A\subseteq [n]$
we denote by $G+ {C}(A)$ the graph obtained from~$G$ by adding edges such
that the subgraph induced on $A$ is a clique. For $n,c\in \mathbb N$ with
$c\in[n]$ and $p\in\mathbb R$ with $0\le p\le 1$ we consider a second
distribution $\ER(n,p,c)$: Pick a random graph $G\in \ER(n, p)$ and a
uniformly random subset $A$ of $[n]$ of size $c$ and plant in $G$ a clique
on $A$, thus getting the graph~$G+ {C}(A)$. The notation~$(G,A)\in
\ER(n,p,c)$ should give the information that the random graph was $G$ and
that the random subset of $[n]$ of size $c$ was $A$.

\begin{theo}\label{thm:indistAC}
Let $k: \mathbb N\to \mathbb R^+$ with $\lim_{n\to \infty} k(n)= \infty$,
and $c:\mathbb N\to \mathbb N$ with $c(n)\le n^{\xi}$ for some $0\le\xi <
1$. Then for all $\AC^0$ circuits $\big(\C_n\big)_{n\in \mathbb N}$,
\[
\lim_{n\to \infty}
 \Pr_{(G,A)\in \ER(n,n^{-1/k(n)},\ c(n))}
 \big[\C_n(G)= \C_n(G+ C(A))\big]
 = 1.
\]
\end{theo}

We first deal with the case  where $k(n)\le \log_2\; n$ for all
sufficiently large $n$. The general case will be reduced to it by standard
techniques from probability theory.
\begin{lem}\label{lem:indistAC}
Let $k: \mathbb N\to \mathbb R^+$ with $k(n)\le \log_2\; n$ for all
sufficiently large $n$ and $\lim_{n\to \infty} k(n)= \infty$, and $c:\mathbb
N\to \mathbb N$ with $c(n)\le n^{\xi}$ for some $0\le\xi < 1$. Then for all
$\AC^0$ circuits $\big(\C_n\big)_{n\in \mathbb N}$,
\[
\lim_{n\to \infty}
 \Pr_{(G,A)\in \ER(n,n^{-1/k(n)},\ c(n))}
 \big[\C_n(G)= \C_n(G+ C(A))\big]
 = 1.
\]
Moreover, the convergence rate is uniform for all $\AC^0$ circuits of a
fixed depth and size.
\end{lem}

\proof Let $(\C_{n})_{n\in \mathbb N}$ be a family of circuits such that for
some $\bar d,t\in \mathbb N$ every $\C_{n}$ has depth at most $\bar d$ and
size bounded by $n^{t}$. In order to apply Lemma~\ref{lem:pccAC}, we set for
$n\in\mathbb N$,
\begin{eqnarray}\label{eq:schl}
S(n) = n^t & \text{and} & d(n)= \bar d.
\end{eqnarray}
By Lemma~\ref{lem:pccAC}, it follows that (recall that $q(n)= n^{-1/k(n)}$)
\begin{equation}\label{eq:Cpcc}
\Pr_{\mu\in \mathscr C^{\ell_{\bar d}(n), q(n)}_n}
 \Big[\text{$\C_n\res{\mu}$ is constant} \Big] = 1- o(1),
\end{equation}
where the $o(1)$ term only depend on $S$, $d$, and $k$, i.e.,
$t$, $\bar d$ and $k$. Furthermore,
\begin{equation}\label{eq:cliquesize}
\ell_{\bar d}(n)= n^{1- \Theta\left(5d(n)\sqrt{(\log_n S(n)
d(n))/k(n)}\right)}=n^{1- o(1)};
\end{equation}
the first equality holds by Lemma~\ref{lem:pccAC} and the second by
\eqref{eq:schl}. The key step consists of the following random process,
which generates $(G,A)\in \ER(n,n^{-1/k(n)},\ c(n))$ from $\mu\in \mathscr
C^{\ell_{\bar d}(n), q(n)}_n$.
\begin{itemize}
\item[(a)] Let $V(G):= [n]$.

\item[(b)] Add edges $e\in \binom{[n]}{2}$ with $\mu(e)=1$ to $E(G)$.

\item[(c)] Recall that $\mu^{-1}(\star)= \binom{U}{2}$, where $U\in
    \binom{[n]}{\ell_{\bar d}(n)}$ was chosen uniformly at random. For
    every $e\in \binom{U}{2}$, add $e$ to $E(G)$ with probability $q(n)$.

\item[(d)] Choose $A\in \binom{U}{c(n)}$ uniformly at random. Note that
    this is possible as $|U|= \ell_{\bar d}(n)= n^{1- o(1)}> n^{\xi}\ge
    c(n)$ for sufficiently large $n$.
\end{itemize}
By (b)--(d), $G$ and $G+C(A)$ contain the same edges from
$\binom{[n]}{2}\setminus\mu^{-1}(\star)$. Thus, by~\eqref{eq:Cpcc},
$\C_n(G)=\C_n(G+C(A))$ with high probability. By (c) and (d), $A$ can be
viewed as being chosen in $\binom{[n]}{c(n)}$ uniformly at random. \proofend

\subsection*{Reduction to small edge probability} We fix the size $c(n)\le
n^{\xi}$ for the planted clique in Theorem~\ref{thm:indistAC}. Assume
$k,k':\mathbb N\to \mathbb R^+$ and $k'(n)\ge k(n)$ for all $n\in \mathbb
N$. We set
\[
p(n):=\frac{n^{-1/k'(n)}- n^{-1/k(n)}}{1- n^{-1/k(n)}}.
\]
Then $0\le p(n)<1$. It is easy to see that for $H\in \ER(n,p(n))$ and $G\in
\ER(n,n^{-1/k(n)})$ the graph $H\cup G$ has the distribution
$\ER(n,n^{-1/k'(n)}) $. Here $H\cup G=\big([n],E(H)\cup E(G)\big)$.
\medskip

Now let $(\C_n)_{n\in \mathbb N}$ be any sequence of circuits of depth $d$
and circuit size $n^t$. For every $H\in \ER(n, p(n))$ one can define a
circuit $\C_n^H$ of depth $d+1$ and size $n^t+n^2$ such that for all graphs
$ G$ with vertex set $[n]$,
\[
\C_n^H(G):= \C_n(H\cup G).
\]
Therefore, we have
\begin{align*}
\Pr &_{(H',A)\in \ER(n,n^{-1/k'(n)},\ c(n) )}
 \big[\C_n(H')= \C_n(H'+ C(A))\big]\\
    & = \Pr_{\substack{H\in \ER(n, p(n)),\\ (G,A)\in \ER(n,n^{-1/k(n)},\ c(n) )}}
       \big[\C_n(H\cup G)= \C_n((H\cup G)+ C(A))\big]\\
    & = \sum_{H_0\in \G(n)} \Pr_{ (G,A)\in \ER(n,n^{-1/k(n)},\ c(n) ) }
         \big[\C^{H_0}_n(G)= \C^{H_0}_n(G+ C(A))\big]
            \cdot \Pr_{H\in \ER(n, p(n))}[H= H_0],
\end{align*}
where $\G(n)$ denotes the set of graphs with vertex set $[n]$. So from the
equality between the first and last term, we see the following.

\begin{prop}\label{prop:red}
Let $c:\mathbb N\to \mathbb N$ with $c(n)\le n^{\xi}$ for some $0\le \xi <
1$ and let $k, k': \mathbb N\to \mathbb R^+$ with $k'(n)\ge k(n)$ for all
$n\in \mathbb N$. If for every $\AC^0$ circuits $(\C_n)_{n\in \mathbb N}$
\[
\lim_{n\to \infty}
 \Pr_{(G,A)\in \ER(n,n^{-1/k(n)},\ c(n) )}
 \big[\C_n(G)= \C_n(G+ C(A))\big]
 = 1
\]
and the convergence rate is uniform for all $\AC^0$ circuits of a fixed
depth and size, then
\[
\lim_{n\to \infty}
 \Pr_{(G,A)\in \ER(n,n^{-1/k'(n)},\ c(n))}
 \big[\C_n(G)= \C_n(G+ C(A))\big]
 = 1.
\]
\end{prop}

Now Theorem~\ref{thm:indistAC} follows immediately from
Lemma~\ref{lem:indistAC} and Proposition~\ref{prop:red}.

\begin{rem}
For a random graph $G\in \ER(n, n^{-1/k(n)})$, the expected size of a maximum clique is $O(k(n))$. Thus if $k(n) = o(c(n))$, to distinguish $G$ and $G+C(A)$ for $(G,A)\in \ER(n,n^{-1/k(n)},\ c(n))$, some constant-depth circuits of size
\[
n^{O(k(n))}
\]
suffice. By a careful inspection of the proof of Theorem~\ref{thm:indistAC}, in particular, the equation~\eqref{eq:cliquesize} in Lemma~\ref{lem:indistAC}, it is easy to see that any constant-depth circuits of size
\[
n^{o(k(n))}
\]
cannot distinguish $G$ and $G+C(A)$.

Furthermore, if the depth of polynomial-size circuits is
\[
o\left(\sqrt{k(n)}\right),
\]
then~\eqref{eq:cliquesize} still holds. Hence, polynomial-size circuits of depth $o(\sqrt{\log\; n})$ cannot distinguish $G$ and $G+C(A)$ for $(G,A)\in \ER\big(n, 1/2, O(\sqrt{n})\big)$.\footnote{For the distribution $(G,A)\in \ER\big(n, 1/2, \Theta(\sqrt{n})\big)$ there are polynomial time algorithms~\cite{alokrisu98} (thus also polynomial-size circuits) which can detect the planted clique $C(A)$ in $G+C(A)$, hence distinguish $G$ and $G+ C(A)$.} These arguments are based on~\eqref{eq:cliquesize}, an equality holding under the hypothesis $k(n)\le \log_2\; n$. Again the general case is reduced by the standard techniques from probability theory used to prove Proposition~\ref{prop:red}.
\end{rem}

\medskip
\subsection{Proof of Theorem~\ref{thm:pgapcliqueparaAC}}\label{sub:app}
Let $\big(\C_{n,k}\big)_{n,k\in \mathbb N}$ be a family of circuits such
that for some function $f:\mathbb N\to \mathbb N$ and $d,c\in \mathbb N$
every $\C_{n,k}$ has depth at most $d$ and size bounded by $f(k)\cdot
n^{c}$. Then we show that there are some $n,k\in \mathbb N$ such that
$\C_{n,k}(G)$ does not decide $\pgapclique \rho$ on instances $(G,k)$ with
$n:= |V(G)|$. Hence, our proof even works for a nonuniform version of
$\paraAC$: We neither assume that the family $(\C_{n,k})$ is computable from
$n$ and $k$ nor that $f$ is computable.

\medskip
We may assume that $f$ is nondecreasing and unbounded. We choose a
nondecreasing and unbounded function $k:\mathbb N\to \mathbb N$ such that
for sufficiently large $n\in \mathbb N$ we have
\begin{equation}\label{eq:k}
2k(n)+1 \le \min\left\{f^{-1}(n), \frac{\sqrt{n}}{\rho(\sqrt{n})}\right\},
\end{equation}
where $f^{-1}(n):= \max \big(\{\ell \mid f(\ell)\le n\}\cup \{0\}\big)$, and
such that $k(n) \le \log_2 n$ for $n\ge 1$. It follows that the circuit
\[
\C:= \C_{n,2k(n)+1}
\]
has size bounded by $S(n):= O(n^{c+1})$, i.e., $\big\{\C_{n,
2k(n)+1}\big\}_{n\in \mathbb N}$ are $\AC^0$-circuits.
%
%

We consider the distribution $(G, A)\in \ER(n, n^{-1/k(n)},
\ceil{\sqrt{n}\,})$. The next claim is easy to verify.

\medskip
\noindent \textit{Claim 1.} $G+C(A)$ contains a clique of size
$\ceil{\sqrt{n}}$, i.e., $\cn(G+C(A))\ge \sqrt{n}$. On the other hand,
$\Pr\big[\cn(G)< 2k(n)+1\big] = 1-o(1)$.
\medskip

\noindent Assume $n$ is sufficiently large, and recall $m\mapsto m/\rho(m)$
is increasing, so~\eqref{eq:k} implies
\[
2k(n)+1 \le \frac{\cn( G+C(A))}{\rho(\cn( G+C(A)))}.
\]
This means that $\big(G+C(A) ,2k(n)+1\big)$ is a yes instance of
$\pgapclique \rho$, while \emph{almost surely} $\big(G ,2k(n)+1\big)$ is a
no instance. Hence, by our assumption on $\big(\C_{n,k}\big)_{n,k\in \mathbb
N}$ and thus on~$\C$,
\begin{align*}
{\hspace{-7mm}} \Pr_{(G, A)\in \ER(n, n^{-1/k(n)}, \ceil{\sqrt{n}})} \big[\C(G+C(A))=1\big]=1,
 \Pr_{(G, A)\in \ER(n, n^{-1/k(n)}, \ceil{\sqrt{n}})} \big[\C(G)=1\big]=o(1).
\end{align*}
But this contradicts Theorem~\ref{thm:indistAC}. \proofend

\medskip
We prove a consequence of Theorem~\ref{thm:pgapcliqueparaAC}. For $t\ge 0$,
$d\ge 1$ we denote by $\Gamma^-_{t,d}$ the subset of subformulas of
$\Gamma_{t,d}$ with only negative literals. Clearly, if $\gamma\in
\Gamma^-_{t,d}$ has a satisfying assignment of Hamming weight $k$, then it
has one of weight $k'$ for every $k'<k$. Denote by $\omega(\gamma)$ the
maximum Hamming weight of assignments satisfying $\gamma$. Then
$\pgapwsat{\rho}(\Gamma^-_{t,d})$ can be defined similarly as $\pgapclique
\rho$.

\begin{prop}
Let $t,d\ge 1$ with $t+d\ge 3$. Then $\pgapwsat{\rho}(\Gamma^-_{t,d}) \notin
\paraAC$.
\end{prop}

\proof Consider the reduction from $\pclique$ to
$\pgapwsat{\rho}(\Gamma_{t,d})$ in the proof of
Corollary~\ref{cor:cliquedswsat}. Clearly $\delta_G\in \Gamma^-_{t,d}$ and
$\delta_G$ is independent of $k$. Thus, the equivalence~\eqref{eq:cliquered}
preserves the approximation ratio. The result then follows immediately.
\proofend

\section{The complexity of $\phalt$}
We already mentioned in the abstract of this article that the complexity of
the parameterized halting problem $\phalt$ is linked to open problems in
computational complexity, descriptive complexity, and proof
theory~\cite{cheflu12}. For example, the membership of $\phalt$ in the parameterized complexity class \emph{uniform XP}  is equivalent to the existence of an almost optimal algorithm for the set of tautologies of propositional logic, or to the fact that a certain logic, presented
in~\cite{gur88}, is a logic for \PTIME. Both statements are conjectured to
be false. The origin of our interest in $\paraAC$ was our hope to get a
lower bound on the complexity of $\phalt$ in terms of $\paraAC$, that is, to
show $\phalt\notin \paraAC$. But also this problem remains open. We know
that $\AC^0$ corresponds to $\FO(<,+, \times)$, first-order logic with an
ordering relation and built-in addition and multiplication. In this section
we prove that $\phalt\notin \para\FO(<,+)$, even $\phalt\notin
\textup{X}\FO(<,+)$, hold unconditionally, to our knowledge the best known
lower bound for the complexity of $\phalt$.

\medskip
Recall that in the paragraph following Lemma~\ref{lem:cc} we defined the
natural ordering $<^{[n]}$ on $[n]$ and the ternary relations $+^{[n]}$ and
$\times^{[n]}$ of addition and multiplication, respectively, on $[n]$. Now
we address the definition of $\X\FO(<,+,\times)$. For this purpose we view
inputs to parameterized problems as structures.

Any string $x\in\Sigma^*$ with $|x|=n$ can be identified with the $\{<, +,
\times, \emph{One}\}$-structure $\langle x\rangle^{<,+,\times}:= ([n],
<^{[n]},+^{[n]}, \times^{[n]}, \emph{One}^{[n]})$. Here $i\in[n]$ is in
$\emph{One}^{[n]}$, the interpretation of the unary relation symbol $
\emph{One}$, if and only if the $i$th bit of $x$ is a `1'. The structures
$\langle x\rangle^{<,+}$ and $\langle x\rangle^{<}$ are reducts of $\langle
x\rangle^{<,+,\times}$ over the vocabularies $\{<, +, \emph{One}\}$ and
$\{<, \emph{One}\}$, respectively.

\begin{defn}\label{def:xfopt}
Let $(Q, \kappa)$ be a parameterized problem. Then $(Q,\kappa)\in
\X\FO(<,+,\times)$ if there is a computable function that assigns to every
$k\in\mathbb N$ a first-order sentence~$\varphi_k$ such that for every
instance~$x$ of $(Q,\kappa)$,
\[
x\in Q \iff \langle x\rangle^{<,+,\times}\models \varphi_{\kappa(x)}.
\]
Analogously, the class $\X\FO(<,+)$ is defined.
\end{defn}

\begin{theo}\label{theo:phxfo}
$\phalt\notin \X\FO(<,+)$.
\end{theo}

\proof For a contradiction we assume that $\phalt\in\X\FO(<,+)$ and show
that then the halting problem for Turing machines would be decidable.

Assume that there is a computable function that assigns to every
$k\in\mathbb N$ a first-order sentence~$ \varphi_k$ such that for every
instance $(1^n,\mathbb M)$,
\[
(1^n,\mathbb M)\in\phalt
 \iff \langle(1^n,\mathbb M)\rangle^{<,+}\models \varphi_{|\mathbb M|}.
\]
Fix $\mathbb M$. There is a first-order interpretation $I$ that for every $
n\in\mathbb N$ defines an isomorphic copy of $\langle(1^n,\mathbb
M)\rangle^{<,+}$ in $([n],<^{[n]},+^{[n]})$: Let $c(n):=|(1^n,\mathbb M)|$
be the length of the string $(1^n,\mathbb M)$. We define the interpretation $I$ stepwise. We choose $ s$ such that $ c(n)\le n^s$. As $ \mathbb M$ is
fixed, in $([n],<^{[n]},+^{[n]})$ we can define the  initial segment of
$[n]^s$ of $ c(n)$ elements, define on it the lexicographical ordering and the relation $ \emph{One}$ such that we get a copy of $\langle(1^n,
\mathbb M)\rangle^{<}$ on this initial segment.  Finally, using $+^{[n]}$, we can define the corresponding built-in addition.

Then, from $\mathbb M$ we can compute $\varphi_{|\mathbb M|}$ and
$\varphi_{|\mathbb M|}^I$ such that
\begin{eqnarray*}
(1^n,\mathbb M)^{<,+}\models \varphi_{|\mathbb M|}
 & \iff & ([n],<^{[n]},+^{[n]})\models \varphi_{|\mathbb M|}^I,
\end{eqnarray*}
and thus,
\begin{equation}\label{eq:phpr}
(1^n,\mathbb M)\in \phalt
 \iff \big([n],<^{[n]},+^{[n]}\big)\models \varphi_{|\mathbb M|}^I.
\end{equation}
By the Ginsburg-Spanier~\cite{ginspa66} improvement of Presburger's Theorem
we know that for $\varphi_{|\mathbb M|}^I$ we may compute $n_0,p_0\in
\mathbb N$ such that for all $n\ge n_0$ we have
\[
\big([n],<^{[n]},+^{[n]}\big)\models \varphi^I_{\mathbb M}
\iff \big([n+p_0],<^{[n+p_0]},+^{[n+p_0]}\big) \models \varphi^I_{\mathbb M}.
\]
By this equivalence and \eqref{eq:phpr} we see that
\[
\text{$\mathbb M$ does not hold on empty input tape}
 \iff \big([n_0],<^{[n_0]},+^{[n_0]}\big)\models \neg\varphi^I_{\mathbb M}.
\]
We can decide the halting problem by checking whether
$\big([n_0],<^{[n_0]},+^{[n_0]}\big) \models \neg\varphi^I_{\mathbb M}$.
\proofend

For the proof it was essential that the function assigning to every
$k\in\mathbb N$ the $\FO(<,+)$-sentence $\varphi_k$ is \emph{computable}.
The class obtained if we drop the requirement of computability in the
definition of $\X\FO(<,+)$ is called $\textup{nonuniform-}\X\FO(<,+)$. We
will see that $\phalt\in \textup{nonuniform-} \X\FO(<,+)$ by the even
stronger statement of Proposition~\ref{pro:phnux}~(1).

We note in passing that by standard modeltheoretic techniques one can show
that the parameterized vertex cover problem, a standard example of a
fixed-parameter tractable problem, is not in the subclass
$\textup{nonuniform-}\X\FO(<)$ of $\textup{nonuniform-}\X\FO(<,+)$. Thus we
get a lower bound for the parameterized complexity of this problem.

We come back to our claim $\phalt\in \textup{nonuniform-} \X\FO(<,+)$. We
even show $\phalt\in \textup{nonuniform-}\para\FO(<,+)$. By definition, a
parameterized problem $(Q,\kappa)$ belongs to the class
$\textup{nonuniform-}\para\FO(<,+)$ (to $\para\FO(<,+)$) if there are a
sentence $\varphi\in \FO(<,+)$ and a (computable) function $\textit{pre}:
\mathbb N \to \Sigma^*$ such that for all $x$,
\[
x\in Q \iff \langle (x, \textit{pre}(\kappa(x))\rangle^{<,+} \models \varphi.
\]
So, in the nonuniform version we allow noncomputable precomputations. Note
that $\para\FO(<,+)\subseteq \X\FO(<,+)$ as the role of the precomputation
\big(in the definition of $\para\FO(<,+)$\big) can be taken over by the
sentences $\varphi_k$ \big(in the definition of $\X\FO(<,+)$\big).

\begin{prop}\label{pro:phnux}
\begin{enumerate}
\item $\phalt\in \textup{nonuniform-}\para\FO(<,+)$.

\item $\phalt\notin \textup{nonuniform-}\para\FO(<)$.
\end{enumerate}
\end{prop}

\proof (1) We look for a first-order sentence $\varphi$ and a function
$\textit{pre}: \mathbb N \to \Sigma^*$ such that for all instances
$(1^n,\mathbb M)$ of $\phalt$,
\begin{equation}\label{eq:endl}
(1^n,\mathbb M)\in \phalt\iff \langle ((1^n,\mathbb M), \textit{pre}(|\mathbb M|)\rangle^{<,+} \models \varphi.
\end{equation}
For a nondeterministic Turing machine $\mathbb M$ define $n_{\mathbb M}$ as
the numbers of steps of a shortest run of~$\mathbb M$ on empty input; set
$n_{\mathbb M}:=\infty$ if every run is infinite. We turn to the definition
of the (noncomputable) precomputation $\textit{pre}:\mathbb N\to\Sigma^*$.
For $k\in \mathbb N$ we enumerate all nondeterministic Turing machines of
length $k$ as
\[
\mathbb M_1, \mathbb M_2, \ldots,\mathbb M_m
\]
and set
\[
\textit{pre}(k):=\$(1^{n_{\mathbb M_1}},\mathbb M_1)\$ \ldots \$(1^{n_{\mathbb M_m}},\mathbb M_m)\$
\] (in a standard way we view $\textit{pre}(k)$ as a string in $\Sigma^* $).
The first-order sentence $\varphi$ satisfying \eqref{eq:endl} expresses that
$ \mathbb M$ is one of the machines $\mathbb M_i$ and that $n_i\le n$. Note
that using addition we can express in first order logic that the substrings
between the points $u_1$ and $u_2$ and between the points $v_1$ and $v_2$
have the same length (by $u_2-u_1=v_2-v_1$) or distinct length, and we also
can express that they coincide.

\medskip
\noindent (2) For a contradiction assume that there is an $\FO(<)$-sentence
$\varphi$ and a function $\textit{pre}: \mathbb N \to \Sigma^*$ such that
for all instances $(1^n,\mathbb M)$ of $\phalt$,
\begin{equation*}
(1^n,\mathbb M)\in \phalt
 \iff \big\langle (1^n,\mathbb M), \textit{pre}(|\mathbb M|)\big\rangle^{<} \models \varphi.
\end{equation*}
We fix a Turing machine $\mathbb M$. Then one easily sees that there is a
first-order interpretation $I=I_{\mathbb M}$ that for all $n\in \mathbb N$
defines a structure
\[
\big\langle(1^n,\mathbb M), \textit{pre}(|\mathbb M|)\big\rangle^{<}
\]
from the structure $([n], <^{[n]})$. Hence, for all $n\in \mathbb N$,
\[
\big\langle (1^n,\mathbb M), \textit{pre}(|\mathbb M|)\big\rangle^{<} \models \varphi
 \iff ([n], <^{[n]})\models \varphi^I.
\]
Let $q$ be the quantifier rank of $\varphi^{I}$. Note that $\varphi^{I}$ is
computable from $\varphi$ and $\mathbb M$. Then we know that for $n,n'>
2^q$,
\[
([n], <^{[n]})\models \varphi^{I}
 \iff ([n'], <^{[n']})\models \varphi^{I}.
\]
So again we can decide the halting problem.

\proofend

Let $\tau$ be a vocabulary which does not contain the reation symbols
$<,+,\times$ and set $\tau_{<,+, \times}:= \tau\cup \{<,+, \times\}$. Recall
that a $\tau_{<,+, \times}$-structure $\str{A}$ \emph{has built-in addition
and built-in multiplication} if $(A, <^{\str{A}}, +^{\str{A}},
\times^{\str{A}})$ is isomorphic to $([|A|],<^{[|A|] }, +^{[|A|] },
\times^{[|A|] })$.

A first-order sentence $\varphi$ of vocabulary $\tau_{<,+, \times}$, shortly
$\varphi\in \FO(<,+,\times)$, is \emph{invariant} (more precisely, \emph{$
<$-invariant}) if for every $\tau$-structure $\str{A}$ and any expansions
$(\str{A},<_1,+_1,\times_1)$ and $(\str{A},<_2,+_2,\times_2)$ of $\str{A}$
to structures with built-in addition and multiplication, we have:
\begin{eqnarray*}
(\str{A},<_1,+_1,\times_1)\models \varphi
 & \iff & (\str{A},<_2,+_2,\times_2)\models \varphi.
\end{eqnarray*}
It should be clear what we mean if we say that a $\varphi\in\FO(<,+)$ or a
$\varphi\in\FO(<)$ is \emph{invariant}.

Along the lines of~\cite[Theorem 10]{cheflu09} one can show:
\begin{prop}\label{pro:pmlo}
Assume that $\phalt\in \X\FO(<,+,\times)$. Let $\tau$ be any vocabulary not
containing the symbols $<$, $+$, and $\times$. Then there is a computable
function $F$ defined on the class of\/ $\FO(<,+,\times)$-sentences of
vocabulary $\tau\cup\{<,+,\times \}$ such that
\begin{itemize}
\item for every $\varphi\in \FO(<,+,\times)$ the sentence $F(\varphi)$ is
    invariant;

\item if $\varphi$ is an invariant $\FO(<,+,\times)$-sentence, then $
    \varphi$ and $F(\varphi)$ are equivalent.

\end{itemize}
Thus, $\{F(\varphi)\mid \varphi \text{\ an invariant $\FO(<,+,\times)$}\}$
is the class of sentences of vocabulary $\tau$ of a logic for the invariant
fragment of\/ $\FO(<,+,\times)$.
\end{prop}
In view of Theorem~\ref{theo:phxfo}, we tried, without success, to show that
for $\FO(<,+)$ there is no computable function $F$ with the properties
mentioned in the preceding result for $\FO(<,+,\times)$, or even to show
that there is no effective enumeration of the invariant sentences
of~$\FO(<,+,\times)$.

\bibliographystyle{plain}
\bibliography{paraFO}
\end{document}